\documentclass[a4paper,11pt]{article}
\pdfoutput=1
\usepackage{jcappub}

\usepackage[english]{babel}

\usepackage{graphicx}
\usepackage{amsmath,amssymb}
\usepackage{hyperref}
\usepackage{braket}
\usepackage{subfigure}
\usepackage{float}
\usepackage[dvipsnames]{xcolor}
\usepackage{mathrsfs}
\usepackage{comment}
\usepackage{multirow}
\usepackage{soul}
\usepackage{mathtools}
\usepackage{sidecap} 
\graphicspath{{Figures/}}
\usepackage[normalem]{ulem}

\title{Proper time path integrals for gravitational waves: an improved wave optics framework}

\author[a]{Ginevra Braga,}
\emailAdd{ginevra.braga@gssi.it}
\author[b]{Alice Garoffolo,}
\emailAdd{aligaro@sas.upenn.edu}
\author[c,d,e]{Angelo Ricciardone,}
\emailAdd{angelo.ricciardone@unipi.it}
\author[e,f,g]{Nicola Bartolo,}
\emailAdd{nicola.bartolo@pd.infn.it}
\author[e,f,g,a]{Sabino Matarrese,}
\emailAdd{sabino.matarrese@pd.infn.it}

\affiliation[a]{Gran Sasso Science Institute, Via F. Crispi 7, 67100 L’Aquila, Italy }

\affiliation[b]{Center for Particle Cosmology, Department of Physics and Astronomy,
University of Pennsylvania 209 S. 33rd St., Philadelphia, PA 19104, USA}

\affiliation[c]{Dipartimento di Fisica “Enrico Fermi”, Università di Pisa, Largo Bruno Pontecorvo 3, Pisa I-56127, Italy}

\affiliation[d]{INFN, Sezione di Pisa,
Largo Bruno Pontecorvo 3, Pisa I-56127, Italy}

\affiliation[e]{Dipartimento di Fisica e Astronomia "Galileo Galilei", Università degli Studi di Padova, via Marzolo 8, I-35131 Padova,Italy}

\affiliation[f]{INFN, Sezione di Padova,
via Marzolo 8, I-35131 Padova, Italy}

\affiliation[g]{INAF-Osservatorio Astronomico di Padova, Vicolo dell’ Osservatorio 5, I-35122 Padova, Italy}

\abstract{
When gravitational waves travel from their source to an observer, they interact with matter structures along their path, causing distinct deformations in their waveforms. 
In this study we introduce a novel theoretical framework for wave optics effects in gravitational lensing, addressing the limitations of existing approaches. We achieve this by incorporating the {\it proper time} technique, typically used in field theory studies, into gravitational lensing.
This approach allows us to extend the standard formalism beyond the eikonal and paraxial approximations, which are traditionally assumed, and to account for polarization effects, which are typically neglected in the literature. We demonstrate that our method provides a robust generalization of conventional approaches, including them as special cases. Our findings enhance our understanding of gravitational wave propagation, which is crucial for accurately interpreting gravitational wave observations and extracting unbiased information about the lenses from the gravitational wave
waveforms.
}

\begin{document}

\maketitle

\section{Introduction}

Gravitational lensing of gravitational waves (GWs) is a rich and complex phenomenon that can provide valuable insights into the matter structures in the Universe. 
One of its intriguing aspects is the possible manifestation of {\it wave optics} (WO) effects: frequency-dependent modulations in the observed waveforms due to interference and diffraction, occurring when the wavelength $\lambda$ of the GW is comparable to the Schwarzschild radius ${\cal R}_s$ of the gravitational lens.
For GWs in the LISA frequency band, these effects are expected to become important for lenses with mass smaller than $ \lesssim 10^8 M_\odot$ and for sources at redshift $z_s \sim 1$, with event rates of $(0.1 - 1.6) \%$ for black-hole binaries in the mass range of $(10^5 - 10^{6.5})\,M_\odot$~\cite{Takahashi:2003ix,Gao:2021sxw}.   
For ground-based interferometers, the possibility of detecting wave-related effects as been assessed in~\cite{Dai:2018enj,Cheung:2020okf}, confirming that these modulations can be detected for events with high enough signal-to-noise ratio.
Given the broadness of the GW frequency spectrum, and the expected increasing in the number of detections with future interferometers~\cite{Reitze:2019iox,Punturo:2010zz,Branchesi:2023mws}, wave optics effects in GW lensing will become an important investigation tool. 
The body of literature exploring the possibility of using WO signatures in GW waveforms for various purposes is growing, given the importance of having more accurate waveforms~\cite{LISAConsortiumWaveformWorkingGroup:2023arg}. These purposes include lens parameter estimation~\cite{Takahashi:2003ix,Caliskan:2022hbu,Tambalo:2022wlm}, uncovering intermediate-mass black holes~\cite{Lai_2018}, investigating the matter power spectrum on small scales~\cite{Tambalo:2022wlm,Savastano:2023spl,Yeung:2021chy,Diego:2019lcd,Guo:2022dre,Jung:2017flg,Fairbairn:2022xln,Urrutia:2024pos}, and constraining the primordial black hole abundance~\cite{Urrutia:2023mtk,Urrutia:2021qak,Sugiyama:2019dgt,Diego:2019rzc,Oguri:2020ldf,GilChoi:2023ahp, Basak:2021ten}.

The current body of literature about wave optics effects is based on the so-called {\it diffraction integral}, derived in the context of gravitational wave's lensing in~\cite{Nakamura1999WaveOI} and further developed in subsequent works~\cite{Nakamura:1997sw,Takahashi:2003ix,Oguri:2020ldf,Takahashi:2004mc,Takahashi:2005sxa,Takahashi:2005ug}.
The formalism is established also for electromagnetic radiation~\cite{Schneider1992,Born_Wolf_Bhatia_Clemmow_Gabor_Stokes_Taylor_Wayman_Wilcock_1999}, and its characteristic trait is to describe WO in a similar language as quantum mechanics. Indeed, in these works the wave equation is recast into a Schr\"odinger-like one, whose solution is given in terms of a path integral where the inverse frequency of the wave, $1/ \omega$, takes the role of $\hbar$~\cite{leung2023wave,goldstein2002classical}. Wave optics phenomena are thus reinterpreted under the light of interference effects between the multiple paths that the wave can take, and, the laws of optics in the high-frequency regime ($\omega \to \infty$), correspond to the classical trajectories found in the $\hbar \to 0$ limit.
This analogy between optics and quantum mechanics is very well known in the literature of electromagnetic waves, and is called {\it Hamilton analogy}~\cite{goldstein2002classical,Masoliver_2009}.
Another avenue to recover the diffraction integral is the one pursued in~\cite{Feldbrugge:2019fjs}, where the authors derive its expression from assigning a particle Hamiltonian, coming from the knowledge of a dispersion relation, and building a path integral description from it. 
Both derivations hinge on the fundamental assumption of being able to describe the wave effectively as a collection of (quantum) particles: the path integral is not performed over all possible field's configuration as in quantum field theory, rather over trajectories as in quantum mechanics.
This fact is reflected also in the type of assumptions the diffraction integral relies on. Two of these are the {\it eikonal} and {\it paraxial} approximations~\cite{Fishman1984,Fishman_2006, Schulman1981TechniquesAA}, which resemble the Wentzel–Kramers–Brillouin (WKB) semiclassical limit of quantum mechanics. 
In order, the eikonal approximation requires that the modifications of the waveform induced by the lens occur on scales larger than the wavelength of the wave. In a similar spirit, the paraxial approximation assumes that any orthogonal displacement from the direction of propagation is negligible within a period. The direct consequence of both of these two assumptions, is to individuate a direction of propagation of the wave, and consequently, to define the wavefronts, at the price of introducing a lower bound in frequency on the validity of the diffraction integral.  Note that the knowledge of wavefronts is equivalent to the assignment of a dispersion relation. 
Given the significant cosmological and astrophysical insights that can be gained from studying the interaction between waves and matter structures, it is important to move beyond the limitations of the diffraction integral and produce a formalism that remains flexible across various length scales, so that we can include a greater range of events in the analysis.

The first aim of this paper is to generalize the treatment of the diffraction integral, going beyond the eikonal and paraxial assumptions. 
We do so by introducing the {\it proper time technique} or {\it worldline formalism} in the field of gravitational wave's lensing. 
This approach, sometimes referred to as {\it first quantised worldline}, was introduced by Feynman and has been developed to formally address the path-integral solution of the Helmholtz equation~\cite{Eve:PI-wavetheory,Fishman1984, Fishman_2006} and it has been used in various fields: ranging from quantum electrodynamics~\cite{Alvarez_1998,space-timeQED,Nambu:1950rs,RevModPhys.58.449}, quantum field theory~\cite{schwinger,Fock:1937dy,Strassler_1992,Bastianelli_2002,Bastianelli:2021nbs,Bastianelli:2023oyz,corradini2021spinning,Bonora:2018uwx,Edwards:2019eby, Edwards:2022dbd} and quantum cosmology \cite{Feldbrugge:2017LQC,Teitelboim:1981ua,Teitelboim:1983fh,Teitelboim:1983fk} under the name of {\it Schwinger proper time}, to light optics with the {\it Feynman-Fradkin} path integral representation~\cite{Schulman1981TechniquesAA,Gutzwiller2004, Gutzwiller_1967, Babington_crossing, Babington_2021,Hannay:pathlinking}, and also in seismology for acoustic wave propagation~\cite{Schlottmann:1999,Palmer1979} .
The proper time technique sees the addition of a new time-like variable, in terms of which the wave equation can be recast {\it exactly} into a Schr\"odinger-like one, with an initial-value constraint, and without the need of any approximation. 
The advantage of this technique, therefore, is to provide an effective description of the fully relativistic fields in terms of associated particles, with their trajectories and associated momenta. The additional temporal parameter is  removed at the end upon integration.
The first main result of this paper is to provide the worldline representation of the gravitational wave's propagator describing their dynamics during a lensing event. This is given as the sum over all possible values of the proper time of particle-like path integrals,  generalizing the diffraction integral of~\cite{Nakamura:1997sw}. 
The reason behind the need of a relativistic quantum mechanical description in terms of particles, rather than fields, lies in the way the high-frequency limit is taken. Working only at the field level, it is not straightforward to see the emergence of the laws of geometric optics (GO) in the high-frequency limit, were the waves can be very well described as a collection of particles.  By assuming eikonal and paraxial approximation {\it a posteriori}, we show that our formalism encompasses the diffraction integral, proving its solidity as its generalization, as well as reproducing the classical laws of geometric optics, such as Snell's laws, the Fermat principle, and the geodesic equations. 
Not only, having a particle-like description for GWs in any optical regime is even more compelling after the first detection of the stochastic gravitational wave background (SGWB) from the pulsar timing arrays collaboration~\cite{NANOGrav:2023gor,EPTA:2023fyk,Reardon:2023gzh,Xu:2023wog,Figueroa:2023zhu}. 
The description of the incoherent superposition of many gravitational waves requires statistical tools, most of which are staged in the phase-space, either through a Boltzmann equation or line-of-sight techniques, where the waves are categorized by their positions and momenta~\cite{Contaldi:2016koz,LISACosmologyWorkingGroup:2022kbp,Bartolo:2019oiq,Bartolo:2019yeu,Ricciardone:2021kel,LISACosmologyWorkingGroup:2022jok,Bertacca:2019fnt,Schulze:2023ich,Malhotra:2022ply, ValbusaDallArmi:2023nqn}, rather than their field configurations.
All of these methods rely on the geodesic equation, a typical feature of the high-frequency regime~\cite{Isaacson1,Isaacson2} where the waves have well defined wavefronts and direction of propagation.
The inclusion of wave optics effects via the introduction of a
collisional term in the Boltzmann equation has been considered
in~\cite{Pizzuti:2022nnj, Cusin:2018avf}.
Our formalism aims at providing a comprehensive
description of the anisotropies of the SGWB, while being truly
agnostic about the optical regime.

The second goal of this work is to address another limitation of the diffraction integral, namely the neglecting of polarization effects. 
Indeed, in this approach GWs are treated as scalar fields, assuming that their amplitude will suffer the largest effects from the lensing events, while its polarization will be unaffected by it. This is a well known high-frequency result~\cite{Isaacson1,Isaacson2}, where the GW's polarization is parallel transported along the paths, and it is assumed to remain valid also in the WO regime. However, it is known that finite size effects can both modify the dynamics of tensor modes and also generate additional polarization content in the GW sector~\cite{Garoffolo:2022usx,Dalang:2021qhu,Cusin:2019rmt}, questioning the validity of the scalar wave assumption. 
geometric optics works in the far field zone ($ \lambda / {\cal R}_s \ll 1 $), where only radiative degrees of freedom are present, and neglects the near zone dynamics which also encompasses non-radiative components of the metric perturbation. Accounting for both of these two sectors could lead to additional interaction channels, producing a richer phenomenology in wave optics phenomena. 
To address the inclusion of polarization effects,  similarly to ~\cite{Pijnenburg:2024btj,Oancea:2022szu,Oancea:2023hgu}, we model the lenses as black holes (BH)  so that we can borrow directly the results of the long-standing field of BH perturbation~\cite{Chandrasekhar:1985kt} in our discussion. 
This research line aims at understanding the behavior of 
BH under an external perturbation predicting, among other results, their quasi normal modes spectra~\cite{Berti:2009kk,Konoplya:2011qq} and the behavior of their tidal deformations encoded in the Love numbers~\cite{Hinderer:2007mb,Binnington:2009bb,Damour:2009vw}.
In the case of binary systems, and especially for extreme mass ratio inspiarls, BH perturbation tools are also employed to compute strong field correction to post-Newtonian results~\cite{Pound:2021qin}. 
The master equation dictating the dynamics of fields perturbations of different spin on a Kerr background is the so-called Teukolsky equation~\cite{Teukolsky:1973ha}, written in terms of Newman-Penrose (NP) scalars~\cite{Newman:1961qr}, proxies for the field perturbations. For instance, when studying $s=2$ metric perturbations, the NP scalars are related to components of the perturbed Weyl tensor, while $s=1$ perturbations to the perturbed Maxwell tensor. At linear order in perturbation theory, NP scalars are also gauge invariant~\cite{Chandrasekhar:1985kt}, so that they are readily related to observables.
In the last part of this work, we will apply the proper time formalism also to Teukolsky equation, showing how to include spin effects into the field of wave optics. 
Our result, therefore, constitute a first step into understanding the role of the interaction with matter on the polarization content of GWs in the context of lensing in the WO regime.

\bigskip
\noindent
This paper is organized as follows: in Section~\ref{sec:nakamura_approximations} we review the derivation of the diffraction integral as done in~\cite{Nakamura:1997sw}, illustrating the role of the eikonal and paraxial approximations. In Section~\ref{sec:ProperTimeScalar} we introduce the proper time technique for a Klein-Gordon field, and derive the worldline representation of the wave's propagator. Section~\ref{sec:HFresults} is dedicated to recover all the standard results of paraxial- and geometric optics, proving that our formalism encompasses them as sub cases. In Section~\ref{sec:perturbative_approach} we set up the perturbative expansion of the Green function. In Section~\ref{sec:Applications} we extend our formalism  to massive scalar fields and we provide explicit computations for a Coulomb-like gravitational potential. Finally, Section~\ref{sec:spin2 for spheric lenses} addresses the inclusion of spin effects for BH lenses. 


\section{The diffraction integral}
\label{sec:nakamura_approximations}
In this section, we review the standard formalism of gravitational lensing of gravitational waves in the wave optics limits, as laid out in the pioneering work of~\cite{Nakamura1999WaveOI} (see~\cite{Nakamura:1997sw,Takahashi:2003ix,Oguri:2020ldf,Takahashi:2004mc,Takahashi:2005sxa,Takahashi:2005ug} for other references),  
bringing into the context of gravitational lensing previous works addressing the analogy between optics and quantum mechanics under the paraxial approximation~\cite{Gloge1969FormalQT,Stoler:1981,guralnik2019new}.

\bigskip
We consider the propagation of a wave through a Universe containing a lens in the Newtonian approximation 
\begin{equation}
\label{eq:background_metric}
    ds^2=-(1+2\alpha U({\bf x}))dt^2+(1-2\alpha U({\bf x}))d{\bf x}^2\,,
\end{equation}
where $U$ is the gravitational potential of the lens, assumed to be static $U=U({\bf x})$. In the expression above we introduced $\alpha$ as the bookkeeping parameter representing the strength of the gravitational potential. In our conventions, therefore, $U$ is of order $\sim 1$ and $\alpha \ll 1$.
\begin{figure}[h!]
    \centering
    \includegraphics[scale=0.5]{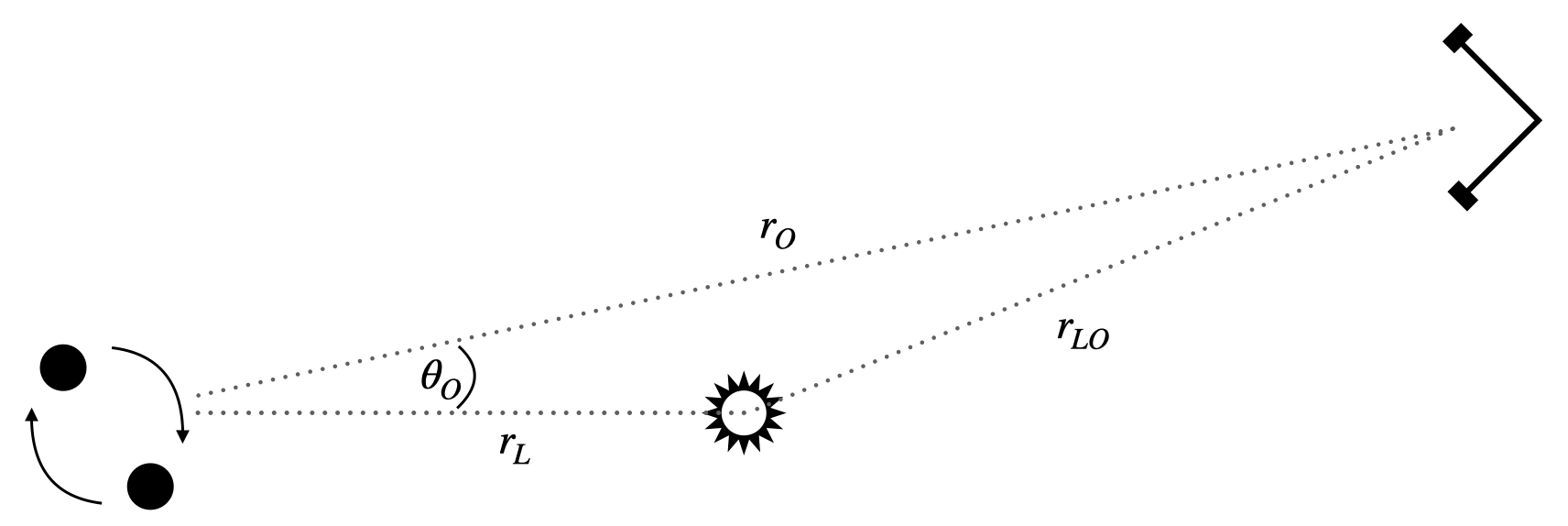}
    \caption{Lensing situation considered.}
   \label{fig:LensngNakamura}
\end{figure}
It is a known fact that, in General Relativity, the polarization content of a GW is parallel transported in the geometric optics limit~\cite{Isaacson1,Isaacson2}. 
In the standard wave optics literature, this result is carried over also in the WO regime, where the GW is treated as a scalar field, namely $\Psi({\bf x},t)$. 
Then, all the physics regarding the interaction between the GWs and the lens is accounted for in the GW's equations of motion on the curved background, namely $\bar \Box \Psi({\bf x},t) = 0$, where the covariant derivative are taken with respect to the metric in Eq.~\eqref{eq:background_metric}. Solving the latter equation is non-trivial due to the non-linearity of the wave operator. One can use the small strength of the gravitational potential ($\alpha \ll 1$) to set up a perturbative expansion, and to study the interaction up to first order in $\alpha$. This approximation is usually called {\it Born approximation}~\cite{Takahashi:2005ug,Takahashi:2005sxa}. 
Since we are considering a static background, we perform a Fourier transform in frequency domain with conventions
\begin{equation}
    \Psi({\bf x},t)=\int \frac{d\omega}{2\pi}\tilde \Psi_\omega({\bf x})e^{-i\omega t}\,,
\end{equation}
and obtain, at first order in $\alpha$, the associated Helmholtz equation 
\begin{equation}
\label{Helmholtz}
    \left[\nabla^2 +\omega^2(1-4\alpha U)\right]\tilde \Psi_\omega({\bf x})=0\,,
\end{equation}
where $\nabla^2\equiv \partial_i \partial_i$. This equation is valid outside the wave's source,
which we assume to be localized at some point ${\bf x}_S$, and generically address it as ${\cal S}({\bf x}_S)$. Eq.~\eqref{Helmholtz} is solved by introducing the amplification factor $F$
\begin{equation}
    F(\mathbf{x}) \equiv \frac{\tilde \Psi_\omega (\mathbf{x})}{\tilde{\Psi}^{NL}_\omega (\mathbf{x})}\,,
\end{equation}
defined as the ratio between the lensed $ \tilde \Psi_\omega $ and unlensed $\tilde{\Psi}^{NL}_\omega$ waves, namely the solution of Eq.~\eqref{Helmholtz} with $U=0$. The Helmholtz equation, then, turns into an equation for the amplification factor 
\begin{equation}
\label{helmh->schrodinger}
   \frac{1}{r^2}\nabla^2_{\boldsymbol \theta} F+\frac{\partial^2 F}{\partial r^2} +2i \omega \frac{\partial}{\partial r} F =  4 \alpha \omega^2 U F \,,
\end{equation}
in spherical coordinates ${\bf x} = \{ r, \theta, \varphi \}$, where we have chosen  ${\tilde{\Psi}^{NL}_\omega (\mathbf{x})} =  e^{-i \omega r}/r$, and $\nabla^2_{\boldsymbol \theta}$ is the Laplacian operator on the 2-D sphere.
It is important to point out that the latter equation has been derived without any assumption on the frequency of the GW. Therefore, in the high-frequency regime, one should recover the standard GO results, while in the opposite limit we expect to have interference and diffraction patterns before reaching the fully diffractive regime~\cite{Schneider1992} where the amplification is suppressed. 
Eq.~\eqref{helmh->schrodinger} allows setting up the previously mentioned analogy between geometric optics (GO) and wave optics (WO) on one side, and classical and quantum mechanics on the other. Indeed, if one is able to remove the second derivative term from it, then the latter becomes a  Schr\"odinger-like equation where the radius takes the role of  the time variable. This way, one could solve Eq.~\eqref{helmh->schrodinger} via a path integral description, which nicely accounts for wave-effects as interference between the multiple propagation paths taken by the GW, in a similar fashion of quantum effects in quantum mechanics.  Throughout the entire paper, we will make such analogy more and more explicit, and exploit it as much as possible to borrow tools from known literature, and bring them in the realm of wave optics effects in gravitational lensing. 
In this review section, we proceed as in~\cite{Nakamura1999WaveOI}, and work out the particle-like path integral solution of  Eq.~\eqref{helmh->schrodinger}. First of all, we have to reduce a second order partial differential equation, to a first order, Schr\"odinger-like, one. 
This can be achieved by neglecting the second order derivative term, with respect to the one proportional to the first order derivative,
\begin{equation}
\label{eq:WKB_Assumption}
    |\partial^2_r F| \ll |2i\omega \partial_r F| \,. 
\end{equation}
If we call ${\cal R}$ the typical variation scale of F (which is related to the typical length scale of the lens' potential) in the direction of motion, then Eq.~\eqref{eq:WKB_Assumption} would imply ${\cal R} \ll \omega$. In \cite{Nakamura1999WaveOI}, the authors call this assumption the {\it eikonal} approximation, as it effectively prescribes a hierarchy between the length scales associated to the GW and those of the lens. It shares also similarities with the non-relativistic limit, except that the role of the time is now assumed by the radial coordinate, and the one of the particle's mass by the frequency.
We also require that the wave's amplitude does not change much in the $\theta$ direction, namely the angle spanning from the line of sight, and that the observer is located at $\theta_{O} \ll 1$ (see Fig.~\ref{fig:LensngNakamura}). This assumption, combined with Eq.~\eqref{eq:WKB_Assumption}, gives us the {\it paraxial approximation}. It allows us to consider the two vector $\boldsymbol{\theta} = \{ \theta, \varphi\}$ as a two dimensional vector on a flat plane and $\sin \theta \approx \theta$ and to treat the system as a $2$-dimensional one, evolved in "time" with the radial coordinate.
As a consequence of Eq.~\eqref{eq:WKB_Assumption},  Eq.~\eqref{helmh->schrodinger} takes the following form
\begin{equation}
\label{schrodi_giappo}
    i\partial_r F=-\frac{1}{2\omega} \partial^2_{ \boldsymbol\theta} F+2 \alpha \omega U F \,,
\end{equation}
where $\partial^2_{ \boldsymbol\theta} = \partial^2 / \partial \theta^2 + \theta^{-1}\partial / \partial \theta + \theta^{-2}\partial^2 / \partial \varphi^2$. \\
We just obtained a Schr\"odinger-like equation, where $r$ plays the role of time and $\omega$ the one of the particle mass. The corresponding Lagrangian, which yields to the classical equation of motion, is given by:
\begin{equation}\label{eq:LagrangianParticle}
    L\left[\boldsymbol{ \theta},\boldsymbol{\dot \theta},r\right]= \left[\frac{r^2}{2}|\boldsymbol{ \dot\theta}|^2-2\alpha U(r,\boldsymbol{\theta} )\right]\,,
\end{equation}
where $\boldsymbol{\dot \theta}=d\boldsymbol\theta/dr$.
We point out that other analogous assumptions have been made in literature in order to reduce the order of the differential equation of $F$. For instance in~\cite{leung2023wave}, the authors introduce the first Fresnel radius $R_F=\sqrt{\lambda D}$, where $D$ is the distance between the source and a given plane, while $\lambda$ is the wavelength of the wave with $\lambda \ll D$. Then, Eq.~\eqref{helmh->schrodinger} turns into a first order differential equation by requiring $R_F \gg \lambda$.
The  {\it diffraction integral} is the path integral solution of Eq.~\eqref{schrodi_giappo}
\begin{equation}\label{eq:PathIntegralNakamura}
    F(\vec r_{O})=\int \mathcal{D}{\boldsymbol \theta}(r) \exp\left\{i \omega \int_0^{r_{O}} dr L\left[{\boldsymbol \theta}(r),\dot {\boldsymbol\theta}(r),r\right]\right\}\,,
\end{equation}
as done in quantum mechanics~\cite{Feynman:100771}, and where ${\bf r}_O$ is the position of the observer.
Therefore, the amplification factor is described in terms of a path integral, where the action weighting each path is built from the particle Lagrangian in Eq.~\eqref{eq:LagrangianParticle}. The diffraction integral in Eq.~\eqref{eq:PathIntegralNakamura} shows manifestly that the inverse frequency $1/ \omega$ is the parameter taking the place of $\hbar$ and formally establishes the analogy between WO effects and quantum mechanical ones. 
The laws of GO, which are valid in the $\omega \to \infty$ limit, are interpreted as the counterparts of the classical limit $\hbar \to 0$, where one or more classical trajectories dominate the path integral solution. These trajectories, namely the {\it rays} along which the wave propagates, are the one that satisfies the Euler-Lagrange equations descending from the Lagrangian in Eq.~\eqref{eq:LagrangianParticle}. Conversely, for smaller values of the GW's frequency, multiple path contribute to the final amplitude, producing interference patterns. 
One can bring Eq.~\eqref{eq:PathIntegralNakamura} in a more familiar form under the {\it thin} lens assumption~\cite{Schneider1992}, namely that the lens extension in the direction of propagation is small compared to the one on the plane orthogonal to it. 
In this case, $ \alpha U(r, \boldsymbol{\theta}) = \frac12 \delta(r - r_L) \hat \psi(\boldsymbol{\theta})$, and the diffraction integral takes the form
\begin{equation}\label{Diffraction_Integral}
    F(\vec r_O) = \int d^2  \boldsymbol{\theta}_L \exp\left\{i\omega \left[ \frac{r_L r_{O}}{2 r_{L O}} |\boldsymbol{\theta}_L - \boldsymbol{\theta}_{O}|^2 - \hat \psi (\boldsymbol{\theta}_L)\right]\right\}\,,
\end{equation}
where $r_L$ is the distance to the lens and $\boldsymbol{\theta}_L$ is the two vector in the lens plane. Note that the path integral has been reduced to a standard 2-dimensional integral under the assumption that the paths contributing the most to the diffraction integral are constant $\boldsymbol{\theta}(t) \approx \boldsymbol{\theta}_L$ (see~\cite{Nakamura:1997sw}).
The integrand of Eq. \eqref{Diffraction_Integral} is the {\it time delay} function
\begin{equation}\label{time delay}
t_d \equiv \frac{r_L r_{O}}{2 r_{L O}} |\boldsymbol{\theta}_L - \boldsymbol{\theta}_{O}|^2 - \hat \psi (\boldsymbol{\theta}_L)\,,
\end{equation}
where $\hat \psi (\boldsymbol{\theta}_L)$ is the deflection potential associated to the lens~\cite{leung2023wave, feldbrugge2020gravitational,Tambalo_2023}. In the high frequency regime, only the stationary points of Eq. \eqref{time delay} contribute to the diffraction integral. These points are obtained by solving the differential equation~\cite{Schneider1992}
\begin{equation}
    \nabla_{{\boldsymbol \theta}_L } t_d=0\,,
\end{equation}
which can be recast in the more familiar form:
\begin{equation}\label{lens equation}
    {\boldsymbol \theta}_O={\boldsymbol \theta}_L-\frac{r_{LO}}{r_L r_O}\nabla_{{\boldsymbol \theta}_L}\hat \psi\,.
\end{equation}
Eq. \eqref{lens equation} is the {\it lens equation}~\cite{Schneider1992, cheung2024probing,Lai_2018, PhysRevD.107.043029}, and determines the position of images that can be produced by a lensing event with an associated strong enough gravitational potential.

\subsection{Field vs particle description}

Before delving into the main body of this work, we aim to address in more details the significance of the approximation in Eq.~\eqref{eq:WKB_Assumption} and the paraxial assumption. 
Firstly let us point out that there is a disagreement in literature concerning what authors means for {\it eikonal} and {\it geometric optics} limits.
Some consider the two description equivalent, while others attribute to the first coherent events and incoherent lensing to the second~\cite{Jow:2022pux,leung2023wave}. 
In the analogy between quantum- and wave- effects, this differentiation divides interference events, occurring only if the trajectories in the path integral "sum" their probability amplitude in phase, from incoherent events where only the classical trajectory remains. 
In our work, we take Eq.~\eqref{eq:WKB_Assumption} as the definition of the eikonal approximation, and reserve the  geometric optics limit, for the saddle point approximation of the path integral. 
Therefore, the eikonal approximation is a half way step:  it does not fully coincide with GO (formally valid when $\omega \to \infty$), but it also requires a lower bound on the possible frequency values. This is because, the interference effects allowed in this intermediate regimes may still be represented in terms of the dynamics of de-phased wavefronts. However, to have wavefronts in the first place, a field must be describable in terms of associated particles (wavefronts define dispersion relations~\cite{Feldbrugge:2019fjs,goldstein2002classical}), regardless of whether their probability amplitudes of transitioning from one spacetime point to another, are summed coherently (as waves), or incoherently (as particles). 

\smallskip
The diffraction integral Eq. \eqref{Diffraction_Integral} also hinges on the paraxial assumption, by considering the propagation occurring manly along a given direction, with slow (compared to the wavelength of the wave) deviations from it~\cite{thorne2021optics}.
Therefore, we can summarize the physical content of the assumptions behind the diffraction integral as the requirements of having a i) slowly changing direction of propagation and ii) wavefronts. 

\smallskip
Our goal is to go beyond the diffraction integral, namely to understand the behavior of GWs when these are treated as {\it fields}, rather than {\it quantum particles}, in the Hamilton analogy.\footnote{We stress that all the phenomena considered in this work are strictly classical, we only rely on $1/\omega \leftrightarrow \hbar$ as an analogy to borrow tools from known literature.} 
We expect  these two descriptions not to be entirely equivalent: transitioning from a relativistic wave equation to a Schr\"odinger-like one (e.g., assuming Eq.~\eqref{eq:WKB_Assumption}) requires certain approximations, excluding possible interesting phenomenology. 
After considering these points, one might argue for solving the Helmholtz equation directly using field theory methods, namely in terms of a path integral over field's configurations rather than trajectories in configuration space. Indeed, the analogy between quantum effects and wave optics extends to the field representation of the wave, and it does not exclusively belong to the particle description.
However, it's important to emphasize the merit and significant advantages of the particle-like description (in terms of trajectories and momenta rather than field variables). Two reasons stand out prominently. 
The first one regards the limit toward the GO regime, 
where the wave's description naturally transitions to one in terms of effective associated particles propagating along the {\it rays}~\cite{Isaacson1,Isaacson2}.
Much of the standard lensing literature in GO is derived from solving the ray equation, thereby describing waves in terms of coordinates and momenta, compatibly with the fact that in the classical limit the uncertainty principle is trivial.  
Hence, having a generalized particle-like description of the field allows a direct high frequency limit toward the well-known geometric optics laws, contrary to a field approach, where this connection is more obscure. 
Secondly, as point out in the Introduction, a particle-like description allows setting up a phase-space where the statistical study of the stochastic gravitational wave background is usually staged.

\section{Proper time path integral for a scalar wave}\label{sec:ProperTimeScalar}

In this section we present our new formalism to generalize the path integral solution of the amplification factor without imposing the eikonal and paraxial approximations of Eq.~\eqref{eq:WKB_Assumption}.
Firstly, we do so in the case of massless scalar fields, and then we generalize to the case of a massive scalar field, and a massless field with spin in the next Sections. 
Our starting point is once again Helmholtz Eq.~\eqref{Helmholtz}, which is completely general. We solve it in terms of the associated Green's function $G_\omega({\bf x}_f,{\bf x}_i)$ satisfying
\begin{equation}
\label{Green_equation}
    \left[\nabla^2 +\omega^2(1-4\alpha U)\right]G_\omega({\bf x}_f,{\bf x}_i)=\delta^{(3)}\left({\bf x}_f-{\bf x}_i\right)\,.
\end{equation}
In order to continue having a particle-like description of the GW, we  have to reduce Eq.~\eqref{Green_equation} to a Schr\"odinger-like equation as well. 
To achieve this description, we introduce the proper time technique in the realm of gravitational lensing,  instead of taking the route illustrated in Section~\ref{sec:nakamura_approximations}.
This is a well-known method in the literature, which prescribes the introduction of a new time-like parameter, $\tau$, called {\it proper time} in a specific way, so that Eq.~\eqref{Green_equation} becomes a first order equation in $\tau$, hence allowing an effective description of the GWs through an associated particle in a higher dimensional spacetime.

\subsection{Adding a new time parameter}
\label{sec:proper_time_parameter}
We apply the proper time method in the case of waves propagating through a spacetime with a lens. The main advantage of this methodology is that it allows a direct interpretation of fields in terms of associated quanta, hence providing a description of the field in terms of paths rather than field configurations.
Here, we propose it for the first time to our knowledge, in a classical gravitational lensing context. Among the references cited in the Introduction, we mainly follow~\cite{Eve:PI-wavetheory, Hannay:pathlinking, Schulman1981TechniquesAA}.

\bigskip
In the proper time method, the system's dynamics is studied in a 5-dimensional spacetime and it is captured by the Green function $\tilde  G_\omega({\bf x}_f,{\bf x}_i, \tau)$. This is related to the 4-dimensional propagator through its definition
\begin{equation}
\label{eq:tau_G}
    G_\omega({\bf x}_f,{\bf x}_i) \equiv-\frac{i}{\omega}\int_0^\infty d\tau\, e^{i\omega \tau} \tilde  G_\omega({\bf x}_f,{\bf x}_i, \tau)\,.
\end{equation}
By inserting this form of $G_\omega$ into its equation of motion, namely Eq.~\eqref{Green_equation}, and performing an integration by part, one finds 
\begin{equation}
   \tilde G_\omega({\bf x}_f,{\bf x}_i,0) + \int_0^\infty d\tau\,e^{i\omega \tau}  \left[\frac{\partial}{\partial \tau}-\frac{i}{\omega}\nabla^2+4i\alpha \omega U \right]\tilde  G_\omega({\bf x}_f,{\bf x}_i,\tau)=\delta^{(3)}\left({\bf x}_f-{\bf x}_i\right)\,.
\end{equation}
Finally, imposing the initial condition $\tilde G_\omega({\bf x}_f,{\bf x}_i,0)=\delta^{(3)}\left({\bf x}_f - {\bf x}_i\right)$, Eq.~\eqref{Green_equation} reduces to a first-order differential equation in terms of the proper time $\tau$
\begin{equation}
\label{Schrodinger_tau}
    \frac{i}{\omega}\frac{\partial}{\partial \tau}\tilde  G_\omega({\bf x}_f,{\bf x}_i,\tau)=-\frac{1}{\omega^2}\nabla^2\tilde  G_\omega({\bf x}_f,{\bf x}_i,\tau)+ V({\bf x}) \tilde  G_\omega({\bf x}_f,{\bf x}_i,\tau)\,,
\end{equation}
where we have defined 
\begin{equation}\label{eq:VandU}
   V({\bf x}) \equiv 4 \alpha  U({\bf x})\,.
\end{equation}
Eq.~\eqref{Schrodinger_tau} is equivalent to a Schr\"odinger equation for a particle moving in a potential $V({\bf x})$. 
Note that we have obtained it without the eikonal and paraxial approximations, hence Eq.~\eqref{Schrodinger_tau}  is completely general and valid for every frequency.
The time evolving parameter is now played by the proper time $\tau$, differently from the diffraction integral where the trajectories where parametrized by the radial coordinate.
As in Section~\ref{sec:nakamura_approximations}, also in Eq.~\eqref{Schrodinger_tau} $1/\omega$ plays the role of $\hbar$, implying that the limit of high frequency $\omega \rightarrow \infty$ can be treated as the semi-classical limit in quantum mechanics, namely the limit of $\hbar \rightarrow 0$.  The fact that the Hamilton analogy is valid also in our formalism, tells us that we expect to recover the GO description~\cite{Isaacson1, Isaacson2} in the high-frequency limit also from Eq.~\eqref{Schrodinger_tau}. In Section~\ref{sec:HFresults} we will show that, by taking the eikonal and paraxial approximation {\it a posteriori}, our formalism is the proper generalization of the WO analysis summarized in Section~\ref{sec:nakamura_approximations}, proving its solidity as a generalization of the standard GO and WO formalism.

\bigskip
Having recovered a Schr\"odinger-like equation, we can now express the associated  solution in the path integral representation~\cite{Eve:PI-wavetheory}
\begin{equation}
\label{Lagrangian_path_integral_Gtilde}
    \tilde  G_\omega({\bf x}_f,{\bf x}_i, \tau)=\int_{{\bf x}(\tau^\prime=0)={\bf x}_i}^{{\bf x}(\tau^\prime=\tau)={\bf x}_f} \mathcal{D}{\bf x}(\tau^\prime) e^{i\omega S}\,,
\end{equation}
where $S$ is interpreted as the associated particle action, taking the form
\begin{equation}
\label{eq:lagrangian}
    S\left[{\bf x},\dot {\bf x}\right]\equiv \int_0^\tau d\tau^\prime L [{\bf x}(\tau'),\dot {\bf x}(\tau')]\,, \qquad \mbox{with} \qquad L=\frac{1}{4}\left(\frac{d{\bf x}(\tau')}{d\tau'}\right)^2-V({\bf x}(\tau'))\,.
\end{equation}
According to Eq.~\eqref{eq:tau_G}, the original propagator is then given by
\begin{equation}
\label{G:path_integral}
    G_\omega({\bf x}_f,{\bf x}_i)=-\frac{i}{\omega}\int_0^\infty \, d\tau\, e^{i\omega \tau}  \int_{{\bf x}(\tau^\prime=0)={\bf x}_i}^{{\bf x}(\tau^\prime=\tau)={\bf x}_f} \mathcal{D}{\bf x}(\tau^\prime) e^{i\omega S}\,.
\end{equation}
Being the proper time $\tau$ integrated over, it does not enter the system Green's function, as  expected since this was introduced as a mathematical device. 
The equivalent form of $G_\omega({\bf x}_f,{\bf x}_i)$ in quantum field theory is known as the {\it worldline representation} of the propagator, and it has been used as an alternative and equivalent avenue to formalize perturbation theory. In this formalism, one focuses on a 1-dimensional field theory, where the particle position $x^\mu$ is seen as a set of four fields living on the one dimensional space of the proper time $\tau$, called {\it worldline}~\cite{Strassler_1992,corradini2021spinning,schwinger}. Namely, one sees the worldline description as the (relativistic) quantum mechanics of the particles that are the quanta of the associated fields in the field theory. Eq.~\eqref{G:path_integral} suggests the following interpretation for the propagator $G_\omega({\bf x}_f,{\bf x}_i)$: it describes the probability for the particle associated to $\tilde \Psi_\omega ({\bf x})$ to propagate from a given initial point located at ${\bf x}_i$, to the final one at ${\bf x}_f$, in a fictitious time $\tau$. The integral over all possible values of $\tau$ means that we have to consider all paths from ${\bf x}_i$ to ${\bf x}_f$ with duration $0<\tau<+\infty$. In other words, it allows for the motion to occur at different ``velocities''.

\bigskip
We also point out that the Green function in Eq.~\eqref{G:path_integral} is defined up to a dimensionless constant, which can be fixed  by physical arguments, such that the propagator in the absence of the potential reduces to the standard Green function of the wave operator. 
Also the boundary conditions are yet fixed. Since we are working in frequency space, one can pick for instance retarded boundary conditions by continuing $\omega \to \omega + i \epsilon$ upon performing the Fourier integral.

\bigskip
Although the Feynman-Fradkin representation of the propagator is an exact form, much of its physics is embodied in the last $\tau-$ integration, and before performing it, it is hard to draw conclusion about the dynamics of  the system~\cite{Fishman1984}.
Because of this, it is sometimes referred to as an {\it indirect} representation. 
On the contrary, direct representations of the Helmholtz propagator, where the integral over  the proper time is removed, are usually approximate in nature. Indeed, to do so one would have to exchange the integration order between the proper time and the path integral, but this is not possible in an exact fashion because of the boundary conditions. 
We shall see  that one can exchange the integration order, or in any case perform the integral over $\tau$, in the high frequency regime, where $\omega$ becomes large~\cite{Garrod_1966,Gutzwiller_1967}. 
Since the aim of this paper is to introduce in the context of GW lensing the proper time technique for the first time, we only show its connection to the existing literature, as its limit toward the high-frequency regime. 
Nonetheless, our propagator encompasses much more physics than that. Because Eq.~\eqref{G:path_integral} is an exact solution, it can be used as starting point to investigate the behavior of GWs near caustics, tunneling and many more effects~\cite{Schulman1981TechniquesAA,Feldbrugge:2019fjs,feldbrugge2020gravitational,Feldbrugge:2023frq}. 
Understanding these phenomena will be the object of future work.

\subsection{Hamiltonian derivation}\label{sec:Hamilton}
The proper time path integral representation of the Green function Eq.~\eqref{G:path_integral} can also be recovered from an Hamiltonian point of view. 
We can identify $G_\omega({\bf x}_f,{\bf x}_i)$, obeying Eq.~\eqref{Green_equation}, as the matrix element
\begin{equation}
\label{eq:Green_function_hilbert_space}
    G_\omega({\bf x}_f,{\bf x}_i)= \braket{{\bf x}_f| \left[ {\bf\nabla}^2+\omega^2(1-4\alpha U)\right]^{-1}|{\bf x}_i}\,,
\end{equation}
taking the form
\begin{equation}\label{eq:Gree_f_PT_hilbert_space}
    G_\omega({\bf x}_f,{\bf x}_i)= -\frac{i}{\omega}\int_0^\infty d\tau\, \braket{{\bf x}_f|e^{i\omega \tau \hat{\mathcal{H}} }|{\bf x}_i}\,,
\end{equation}
where the particle Hamiltonian is given by
\begin{equation}\label{particle_Hamiltonian}
    \hat{\mathcal{H}}
     \equiv -p^2 + (1 - 4 \alpha U)\,,
\end{equation}
with ${\bf p} \equiv i \omega^{-1} \boldsymbol{\nabla}$ and $p^2 = {\bf p} \cdot {\bf p}$.
As usual (see e.g.~\cite{Schneider1992}), the propagation of a wave through the background curved by the lens can also be seen as the problem of a wave propagating through a medium with refractive index
\begin{equation}\label{refractive_index}
    n^2 \equiv  1-4\alpha U \,, \qquad \to \qquad n \simeq 1 - 2 \alpha U 
\end{equation}
hence the particle Hamiltonian would be given by $\hat{\mathcal{H}} = -p^2 + n^2 $.
Eq.~\eqref{eq:Gree_f_PT_hilbert_space} clearly shows that the proper time allows for a particle-like description of a field. Indeed, the matrix element inside the integral is particle propagator, given in terms of a particle path integral, while the field propagator is $G_\omega({\bf x}_f,{\bf x}_i)$, expressed as an integral over the proper time. 
Despite that a field description and a particle-one are not equivalent, one can achieve a particle-like description of the waves with the aid of the proper time. 
The Hamiltonian derivation of the field propagator shows that one can introduce the additional time-like variable exploiting first quantization tools and not just embedding techniques as in Eq.~\eqref{eq:tau_G}. 
Moreover, it nicely shows that the gravitational potential $U$ can be interpreted as the {\it self-energy}, dressing the free propagator in momentum space, and supports its identification with the index of refraction. 
In Section~\ref{sec:HigherOrderPTDyson} we will also show that it satisfies Dyson equations.

Note that this Hamiltonian derivation of the worldline representation of the propagator is also used in~\cite{Feldbrugge:2019fjs} to derive the diffraction integral. However, the proper time is fixed by considering fixed average energy configurations.
Indeed, the Green function in Eq.~\eqref{eq:Green_function_hilbert_space} can be interpreted as the {\it fixed energy propagator} \cite{Schulman1981TechniquesAA}.

\section{The high-frequency limit}\label{sec:HFresults}

In this section we prove that the formalism we set up is able to recover both the results of WO literature (as described in Section~\ref{sec:nakamura_approximations}) and the GO laws of propagation by taking the high-frequency limit of Eq.~\eqref{G:path_integral}.

\subsection{Geometric Optics: the ray equation}
In the limit in which $\omega \to \infty$,  the complex exponential in Eq.~\eqref{G:path_integral} becomes highly oscillating and the main contribution to the path integral comes from the paths that are close to the minima of the phase. 
Therefore, the rays of the GO description emerge naturally in the high frequency limit. \\
We start by considering the total action
\begin{equation}
    W[{\bf x}, \, \dot{\bf x},\,  \tau] \equiv \tau + \int^\tau_0 d \tau' L [{\bf x}(\tau'), \dot{\bf x}(\tau')]\,,
\end{equation}
which is a function of both the proper time and the trajectories. The function $W$ is also known as the {\it Hamilton characteristic function}~\cite{goldstein2002classical, Masoliver_2009}.
The equation of motion of $\tau$ gives the constraint 
\begin{equation}
\label{Constraint_refractive}
   \frac{\partial W}{\partial \tau} =  1-\left[\frac{1}{4}\left(\frac{\text d {\bf x}(\tau')}{\text d \tau'}\right)^2+V({\bf x}(\tau'))\right]=0\,,
\end{equation}
selecting a specific value of $\tau$ as a function of the trajectory and the boundary conditions, which we call $\tau_{cl}$ (this enters in the boundary condition of the trajectory). Considering, for instance, a free particle, the only path must satisfy
\begin{align}\label{eq:freeEQ}
    \ddot {\bf x}_{free}(\tau') &= 0\,, \qquad 1 - \frac{1}{4} |\dot {\bf x}_{free}|^2 = 0\,,
\end{align}
with boundary conditions ${\bf x}_{free}(\tau' = 0 ) = {\bf x}_i$ and ${\bf x}_{free}(\tau' = \tau ) = {\bf x}_f$, implying $\dot {\bf x}_{free} \equiv  ({\bf x}_f - {\bf x}_i)/ \tau$, and 
\begin{equation}\label{eq:tauEFree}
    \tau_{free} = \frac{|{\bf x}_f - {\bf x}_i|}{2}\,.
\end{equation}

\begin{figure}[H]
    \centering
    \includegraphics[width=0.7\textwidth]{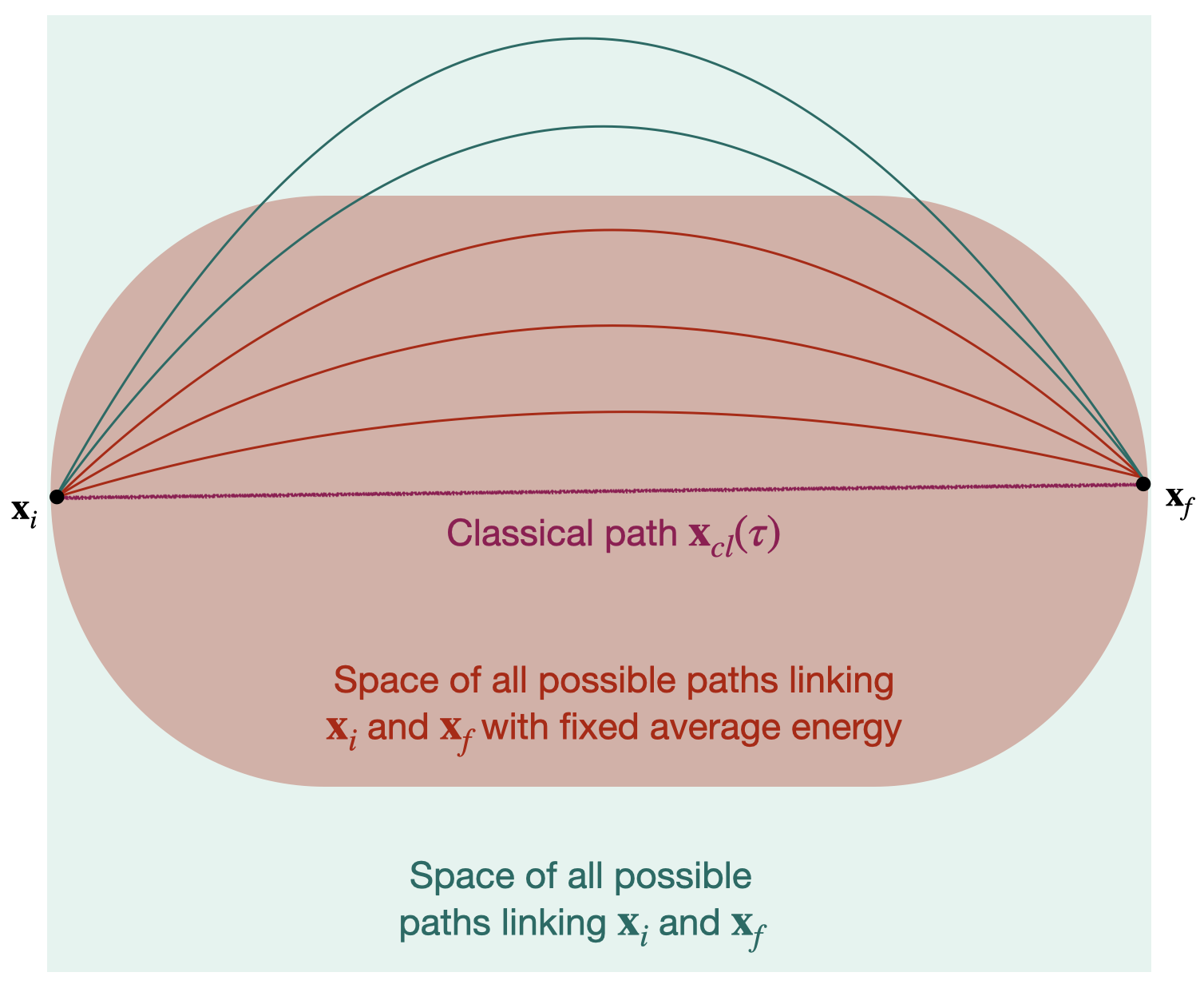}
    \caption{Among all the possible paths (green region), the constraint in Eq.~\eqref{Constraint_refractive} selects only those paths connecting ${\bf x}_i$ and ${\bf x}_f$ with fixed average energy. The classical path satisfies $\delta \hat S / \delta x = 0$.}
   \label{fig:disegnino}
\end{figure}

Eq.~\eqref{Constraint_refractive} is related to the conservation of energy.
One might have thought that computing the variation of the phase with respect to the proper time $\tau$, would have lead to a constraint on the associated Lagrangian $L$ in Eq.~\eqref{eq:lagrangian}.
However, this is not the case as in changing $\tau$, one also changes the path along which the action $S$ is evaluated. Specifically, it can be shown~\cite{Feynman:100771, Schulman1981TechniquesAA} that
\begin{equation}
    \frac{\partial W}{\partial \tau}=- \hat {\cal H}\,,
\end{equation}
where $\hat {\cal H}$ is the associated Hamiltonian of  Eq.~\eqref{particle_Hamiltonian}. Indeed, one can derive Eq.~\eqref{Constraint_refractive} also from $\hat {\cal H}=0$ by plugging the expression of the velocity $\dot {\bf x}$, related to the momenta through Hamilton equations $\dot {\bf x} = \partial \hat {\cal H} / \partial {\bf p}$, in $\hat {\cal H}$.
Therefore, the Hamiltonian constraint above selects those trajectories where the total energy is conserved (see Figure~\ref{fig:disegnino}), among all the possible paths.
The fact that it is equivalent to the vanishing of the Hamiltonian, can be also interpreted as the assignment of a dispersion relation, which defines the wavefronts, since $\hat {\cal H} = 0$ implies $p^2 = n^2$. 
This is also supported by the fact that Eq.~\eqref{Constraint_refractive} is the Hamilton-Jacobi equation for the particle action $S$.
We can also give an interpretation of this result in light of the eikonal approximation described in Section~\ref{sec:nakamura_approximations}. The diffraction integral has among its basic assumptions the one of transforming a second order (in $r$) partial differential equation, to a first order one. We also commented that such a procedure is similar to taking the non-relativistic limit from a Klein-Gordon to a Schr\"odinger one. Hence, we can read the fact that removing the proper time fixes one particular $\tau$ parametrization under the light of the non-relativistic approximation. 

Note that the constraint in Eq.~\eqref{Constraint_refractive}, is satisfied on the classical trajectory up to a constant. The Euler-Lagrange equations of the particle's action~\eqref{eq:lagrangian}, pick out the path
\begin{equation}\label{eq:EulerLagrange}
  \ddot {\bf x}_{cl}  = - 2  \boldsymbol{\nabla} V [{\bf x}_{cl}] \,,
\end{equation}
and the time derivative of the Hamiltonian constraint
\begin{align}\label{eq:SecondTauDerivativeW}
   \frac{\partial^2 W}{\partial \tau^2} =  \dot {\bf x} \cdot \left[ \frac{\ddot {\bf x}}{2}  + \boldsymbol{\nabla} V [{\bf x}]\right]\,,
\end{align}
vanishes on the classical trajectory in virtue of the Euler-Lagrange Eq.~\eqref{eq:EulerLagrange}. Therefore, on ${\bf x}_{cl}$ the constraint Eq.~\eqref{Constraint_refractive} is already implemented up to a constant. This is the usual result stating that only for the on-shell motion energy is well defined (begin constant in time). 
However, it also means that the the knowledge of ${\bf x}_{cl}$ does not fix $\tau_{cl}$ because the constraint equation is only recovered up to an arbitrary constant.

\bigskip
We can rewrite the total action W in a more familiar form by chaining the path's parametrization from the proper time to the path's length $dl = \sqrt{\dot {\bf x} \cdot \dot {\bf x}} \, d \tau$, leading to 
\begin{equation}
\label{fictituous action}
    W=\int_0^{l_{cl}} dl^\prime \sqrt{1-4\alpha U({\bf x}(l'))} = \int_0^{l_{cl}} dl^\prime n({\bf x}(l'))\,,
\end{equation}
where $n$ is the refractive index defined in Eq.~\eqref{refractive_index} and where we have used the constraint in Eq.~\eqref{Constraint_refractive}.
We can use this result to find the equations of the trajectories, namely the rays of GO. These follow from the variational principle applied to  Eq.~\eqref{fictituous action}
\begin{equation}\label{eq:FermatPrinciple}
    \delta \left(\int_0^{l_{cl}} dl^\prime n({\bf x}(l^\prime))\right)=0\,,
\end{equation}
leading to the known Fermat principle~\cite{Schneider1992,Born_Wolf_Bhatia_Clemmow_Gabor_Stokes_Taylor_Wayman_Wilcock_1999,goldstein2002classical}, a fundamental part of optics in the high frequency regime. 
Note that the knowledge of a ray, gives automatically the phase space variables: a unique trajectory has a well defined position (${\bf x}^\mu$) and momentum ($\text d {\bf x}^\mu / \text d l$) vectors at all points.

\subsection{WKB form of the propagator}
The results just found concern the stationary points of the path integral action.  We can do better and study the contributions of the first non-null fluctuations around the classical configuration, which will dictate how the field's amplitude changes during the propagation. 
As we will show, these contributions are not suppressed by a factor $1 / \omega$ in the final results, they are an integrating part of the GO result.
Following~\cite{Schulman1981TechniquesAA,Gutzwiller_1967,Feldbrugge:2023frq}, we start the analysis from the unconstrained particle action $S$ (rather than $W$) in Eq.~\eqref{eq:lagrangian} and perform a saddle-point approximation first in the trajectories, and then in the proper time. The procedure leads to
\begin{align}\label{eq:GWKB}
    G^{\text{\tiny{WKB}}}_\omega({\bf x}_f, {\bf x}_i) \approx {\cal N} \: \sum_{IJ} \sqrt{ - \frac{\mbox{det} [\partial^2 S_{cl}/ \partial x^a_i \partial x^b_f]}{\partial^2 S_{cl}/\partial \tau^2} \Bigg|_{I}} \: e^{i \omega W[{\bf x}^I_{cl}, \tau^{IJ}_{cl}] - i \pi n^I / 2}  \,,
\end{align}
where we have used the Van Vleck-Morette determinant~\cite{Schulman1981TechniquesAA}. 
The sum over $I$, corresponds to the sum over different classical configurations, i.e., different images, if Eq.~\eqref{eq:EulerLagrange} has multiple solutions, while the one over $J$ accounts for possible different values of proper time fixed by the constraint Eq.~\eqref{Constraint_refractive} for each image.
${\cal N}$ is a dimensionless normalization constant that has to be fixed,  and the factor $n^I$ is the number of negative eigenvalues of the Hessian matrix of the image $I$, also related to the number of focal points lying on the classical trajectory~\cite{Schulman1981TechniquesAA}, as stated by Morse theory. 
The amplitude of the WKB propagator can, in principle, diverge. These can occur on caustics (infinities of the Van Vleck-Morette determinant) or at the zeros of the denominator. 
Concerning the first, these are lines of infinite magnification, signaling the breaking of the saddle-point approximation when two classical trajectories merge into each other.  On the other hand, the zeros of $\partial^2 S_{cl} / \partial \tau^2$ are found in correspondence of the turning points of the trajectory, namely when $\dot {\bf x} = 0$ and the system changes direction of motion~\cite{Schulman1981TechniquesAA}.

As an example, we compute the square root in Eq.~\eqref{eq:GWKB} in the absence of the potential, and show that this leads to the decay with distance $\propto 1 / |{\bf x}_f - {\bf x}_i|$ typical of the free propagation. 
Indeed, if one assumes a GW source localized at a certain spacetime point ${\bf x}_s$, then the field solution takes the form 
\begin{equation}
    \Psi_\omega ({\bf x} ) =  {\cal S}  ({\bf x}_s) \, G^{\text{\tiny{WKB}}}_\omega({\bf x}, {\bf x}_s) \,,
\end{equation}
where ${\cal S}  ({\bf x}_s)$ is the value of the source's amplitude at its location. Hence, the decay with distance typical of freely propagating fields must be contained in the Green function.
The first ingredient we need is the classical trajectory. 
In the absence of the potential, this satisfies Eq.~\eqref{eq:freeEQ}, and the velocity is given by $\dot {\bf x}_{cl} \equiv  ({\bf x}_f - {\bf x}_i)/ \tau$. 
Therefore
\begin{align}
    S_{cl}[\tau] = \frac{|{\bf x}_f - {\bf x}_i|^2}{4 \tau}\,.
\end{align}
The derivatives entering the square root in Eq.~\eqref{eq:GWKB} are then easily computed
\begin{align}
    \frac{\partial^2 S_{cl}}{\partial x^a_f \partial x^b_i} = - \frac{\delta_{ab}}{2 \tau} \,, \qquad 
    \frac{\partial^2 S_{cl}}{\partial \tau^2} = \frac{|{\bf x}_f - {\bf x}_i |^2}{2 \tau^3} \,,
\end{align}
giving
\begin{align}
    G^{\text{\tiny{WKB}}}_\omega({\bf x}_f, {\bf x}_i) = -\frac{1}{4 \pi} \frac{e^{i \omega|{\bf x}_f - {\bf x}_i |} }{ |{\bf x}_f - {\bf x}_i | } \,,
\end{align}
since the Morse index is null in this case, and where we have used Eq~\eqref{eq:tauEFree} to compute the phase.
We have recovered the free propagator (see later Eq.~\eqref{eq:FreeProp} for its explicit computation), and fixed the normalization constant accordingly, choosing ${\cal N} = - 1 / 2 \pi$. 
The same computation can be performed again accounting also for the gravitational potential, if one knows how to solve the equations of motion of the classical trajectory. 
To keep the potential generic, here we take a perturbative approach considering that its effects on the trajectory are small. Recalling that $V = 4 \alpha U$, we use the smallness of $\alpha$ and solve Eq.~\eqref{eq:EulerLagrange} as ${\bf x}_{cl}(\tau') = {\bf x}_{free}(\tau') + \alpha {\bf q}(\tau')$, where $\ddot {\bf x}_{free} = 0$ while 
\begin{equation}
    \frac{d^2 \, \bar {\bf q}(\tau')}{ d {\tau'}^2} = - 2 \boldsymbol{\nabla} V[\bar {\bf x}_{free}(\tau')]\,,
\end{equation}
with boundary conditions  ${\bf q} (\tau' = \tau) = {\bf q} (\tau'= 0) = 0$.
Therefore the action evaluated on classical path is
\begin{align}
        S_{cl}[\tau] &= \int^\tau_0 d \tau' \, \left[ \frac{1}{4} \left( \frac{d  {\bf x}_{cl}}{d {\tau'}}\right)^2 - V[{\bf x}_{cl}(\tau')]\right] = \nonumber \\
        &= \frac{|{\bf x}_f - {\bf x}_i |^2}{4 \tau}  + \alpha \int^\tau_0 d \tau' \, \left[ \frac{{\bf x}_f - {\bf x}_i}{2 \tau} \cdot \frac{d \bar {\bf q}}{d {\tau'}} - 4 U[\bar {\bf x}_{free}(\tau')]\right]  = \nonumber \\
        &= \frac{|{\bf x}_f - {\bf x}_i |^2}{4 \tau} + \alpha {\cal V}\,,
\end{align}
where in the last line we defined ${\cal V}$ collecting all together the first order contributions.
From this expression, we can take again the second derivatives of the action evaluated on the classical trajectory with respect to the initial and final points, and the proper time
\begin{align}
     \frac{\partial^2 S_{cl}[\tau]}{\partial x^a_f \partial x^b_i} = - \frac{\delta_{ab}}{2 \tau} + \alpha \frac{\partial^2 {\cal V}}{\partial x^a_f \partial x^b_i} \,, \qquad 
     \frac{\partial^2 S_{cl}[\tau]}{\partial \tau^2} = \frac{|{\bf x}_f - {\bf x}_i |^2}{2 \tau^3} + \alpha \frac{\partial^2 {\cal V}}{\partial \tau^2}
\end{align}
yielding the WKB form of the propagator
\begin{align}\label{eq:WKBGreen1}
    G^{\text{\tiny{WKB}}}_\omega({\bf x}_f, {\bf x}_i) \approx  -\frac{1}{4 \pi} \:  \frac{e^{i \omega W[{\bf x}_{cl}, \tau_{cl}]}}{|{\bf x}_f - {\bf x}_i|} \: \left[ 1 - \alpha \frac{|{\bf x}_f - {\bf x}_i|}{2} \left( \mbox{Tr} \left[ \frac{\partial^2 {\cal V}}{\partial x^a_f \partial x^b_i}\right] + \frac14 \frac{\partial^2 {\cal V}}{\partial \tau^2} \right)\Bigg|_{\tau_{cl}} \right]\,,
\end{align}
where we have fixed the normalization constant in the limit $\alpha \to 0$ using the free propagator.
Since we performed the expansion in $\alpha \ll 1$, this result shows the decay with distance typical of the wave's propagation, at zero-th order in the parameter. 
The presence of the lens generates both a modification of the amplitude (the second term in the parenthesis) and in the phase ($W[{\bf x}_{cl}, \tau_{cl}]$ has a term ${\cal O}(\alpha)$). Both of these two terms are frequency independent, as it should since we are considering the high-frequency limit and the GO laws are achromatic.
Note also that the assumption $\alpha \ll 1$ implies that there is only one classical trajectory, reason why we removed the summation over the different images.

\subsection{Recovering the diffraction integral under the paraxial approximation}
\label{paraxial_approach}

In this section we want to show that Eq.~\eqref{G:path_integral} reduces to the WO literature presented in Section~\ref{sec:nakamura_approximations} starting from our propagator given in the worldline description.
Given all that we discussed, few points are clear: we should recover the diffraction integral by imposing the paraxial and eikonal approximations on our Green function and, in doing so, the integral over the proper time should be removed.
Indeed, the main difference between our approach and the diffraction integral is the presence of the integral over all possible values of the proper time. If one was able to remove this last step, then the two description should match. 
In general, though, it is not possible to exchange the integration order between the proper time and path integrals and perform the former first, because the latter has integration boundaries which depend on $\tau$. 
Nonetheless, exchanging the two integrals can be done in the high-frequency limit~\cite{Fishman1984}, as shown by the  {\it Feynman-Garrod} representation of the propagator~\cite{Fishman1984,Fishman_2006,DeWittMorette1979PathII,Gutzwiller_1967}.
To achieve it, we assume to pass the proper time integral through and perform a half-step compared to the derivation of the WKB form of the propagator, fixing the value of the proper time, while leaving free to vary the trajectories.
This approach can be carried out in two ways: performing a saddle point approximation considering variations in $\tau$ around its classical configuration, or by strictly imposing the vanishing of the Hamiltonian, hence requiring that Eq.~\eqref{Constraint_refractive} holds exactly and simply evaluating the integral in $\tau_{cl}$. Practically, both of the approaches imply a series in $\delta \tau = \tau - \tau_{cl}$ of the phase $e^{i \omega W}$, and in the first case the series is truncated at second order, while in the second case at first order.
We illustrate both of the two options, showing that the first leads to the Feynman-Garrod form of the propagator while the second one to the diffraction integral in the paraxial limit. Note that this derivation is similar to the one displayed in~\cite{feldbrugge2020gravitational} where the  diffraction integral is found starting from the dispersion relation.

\subsubsection{The diffraction integral}
We start with the approach that leads to the diffraction integral, namely requiring that the Hamiltonian constraint is valid exactly. 
Therefore we simply evaluate the integral in $\tau$ as a suitable normalization (taking care of the dimensions) times the path integral evaluated in the classical configuration $\tau_{cl}$.
Enforcing Eq.~\eqref{Constraint_refractive} and writing the path integral in terms of the arc length, instead of the proper time, leads us to
\begin{equation}
    G_\omega({\bf x}_f, {\bf x}_i) \approx  {\cal N} \int_{{\bf x}(l^\prime=0)={\bf x}_i}^{{\bf x}(l^\prime=l_{cl})={\bf x}_f} \mathcal{D}{\bf x}(l^\prime) \:e^{i\omega \, \int^{l_{cl}}_0 d l' n({\bf x}(l')) }\,,
\end{equation}
showing that the action weighting the paths is given by the refractive index, as it known in optics~\cite{Feldbrugge:2019fjs,Schneider1992,Born_Wolf_Bhatia_Clemmow_Gabor_Stokes_Taylor_Wayman_Wilcock_1999,ThorneBlandford}.
The normalization factor  $\mathcal{N}$ will be subsequently determined by requiring that in the limit $\omega R_s \to 0$ we find the free propagator of Eq. \eqref{eq:FreeProp}. Hence we expect a behaviour like ${\cal N} \propto 1 / |{\bf x}_f - {\bf x}_i|$. 
Having removed the integral over the proper time, we proceed in evaluating the Green function under the paraxial approximation by considering waves propagating along a main direction, with small deviations from it. In full generality, we choose the propagation direction to be along the $z$ axis and that the lens is thin. 
We redefine the spatial coordinates as ${\bf x} = (\vec x,z)$, where $\vec x$ is the 2-dimensional vector $\vec x= (x^1,x^2)$.
The constrained total action receives two contributions 
\begin{equation}
    W = \int_0^{l_{cl}} dl^\prime \, n({\bf x}(l^\prime))= \int_0^{l_{cl}} dl^\prime \,  [1 -2 \alpha U({\bf x}(l^\prime))] = l_{cl}-2\alpha \int_0^l dl^\prime U({\bf x}(l^\prime))\,, 
\end{equation}
where  the first term represents the geometric time delay, $t_{geo}$, while the second one is the gravitational time delay $t_{grav}$ experienced by the paths immersed in the gravitational field of the lens.
The geometric time delay can be integrated using the Pythagorean theorem in the paraxial approximation. We implement this by requiring that the rays bend instantly as they hit the thin lens plane, and the horizontal displacement has to be much smaller than the distances $r_{O}$ and $r_{L}$ separating the observer from the lens and the source from the lens (see Figure~\ref{fig:LensngNakamura}).
After some manipulations, we obtain
\begin{equation}
\label{eq:t_geometric}
    t_{geom}=l_{cl}=\frac{1}{2}\frac{r_{O}}{r_{LO}r_{L}}\left(\vec x-\vec \mu\right)^2\,, \qquad \mbox{where} \qquad \vec \mu\equiv \frac{\vec x_{O}r_{L}+\vec x_S r_{LO}}{r_{O}}
\end{equation}
where $\vec x_{O}$ and $\vec x_S$ are the 2-D vectors describing the observer and the source position in the lens' plane, as in~\cite{Feldbrugge:2019fjs}.
As for the gravitational time delay $t_{grav}$, the thin lens approximation allows us to approximate the integral by collapsing the gravitational field over the $z$ axis. In this way we can re-express $t_{grav}$ as a function of the impact parameter $\vec x(l)$ in the following way
\begin{equation}
\label{eq:t_Grav}
    t_{grav}=-2\alpha \int_0^{l_{cl}} dl^\prime U({\bf x}(l^\prime)) = -2\alpha \int U(\vec x+ z\hat z) dz \equiv - \hat \psi(\vec x)\,.
\end{equation}
As explained in~\cite{Nakamura1999WaveOI}, the result of the thin-lens approximation is that the paths that contribute the most to the phase integral are well approximated by $\vec{x}(l) = \vec{x}_L$, namely by a constant vector. In this case the path integral collapses to a 2D integral on the planes orthogonal to the $z$ direction
\begin{equation}
     G_\omega({\bf x}_O,{\bf x}_S) = {\cal N} \int_{\vec x_S}^{{\vec x}_O} d\vec x\, \exp\left\{i\omega \left[\frac{1}{2}\frac{r_{O}}{r_{LO}r_{L}}\left(\vec x-\vec \mu\right)^2-\hat \psi(\vec x)\right]\right\}\,,
\end{equation}
where we have considered the source located in ${\bf x}_i$ and the observer in ${\bf x}_f$. 
This result matches the diffraction integral found in~\cite{Feldbrugge:2019fjs}, and it is equivalent to the one described in Section~\ref{sec:nakamura_approximations}. Thus we have proved that our formalism encompasses the current WO literature in the eikonal and paraxial limits.

\subsubsection{The Feynman-Garrod propagator}

Instead of enforcing the Hamiltonian constraint strictly, one can also expand the total phase up to second order in $\tau - \tau_{cl}$, and perform the Gaussian integral over the proper time.
This requires to exchange the order of the proper time and the path integral, procedure which is allowed in the high frequency regime~\cite{Garrod_1966,Fishman_2006,DeWittMorette1979PathII,Gutzwiller_1967}. Doing so, leads to the Feynman-Garrod representation of the propagator
\begin{align}\label{eq:FeynmanGarrod}
    G_\omega({\bf x}_f, {\bf x}_i) \approx {\cal N} \int_{{\bf x}(\tau^\prime=0)={\bf x}_i}^{{\bf x}(\tau^\prime=\tau_{cl})={\bf x}_f} \mathcal{D}{\bf x}(\tau^\prime) \frac{e^{i\omega \, W [{\bf x}, \, \dot{\bf x}, \, \tau_{cl}]  } }{\sqrt{[ \partial^2 W[{\bf x}, \, \dot{\bf x}, \, \tau_{cl}]/\partial \tau^2]_{\tau_{cl}}}}\,,
\end{align}
where the normalization still needs to be fixed suitably.
The square root at the denominator comes from the Gaussian integral and it can develop poles, corresponding points where the approximation performed breaks down. Indeed, the enforcing of the Hamiltonian constraint leads to the existence of caustics~\cite{guralnik2019new, thorne2021optics,Schulman1981TechniquesAA} where the saddle point approximation breaks down, as it is unable to account for nontrivial wave contributions.
As it can be understood from Eq.~\eqref{eq:SecondTauDerivativeW}, the zeros of the denominator in Eq.~\eqref{eq:FeynmanGarrod} are found in correspondence of:  the classical trajectories, turning points where $\dot {\bf x} = 0$, or points where the velocity becomes orthogonal to the acceleration and the force exerted on the particle.
Regarding the first case, from the WKB analysis carried out previously, we know that this divergence is not there: when the classical path dominates the path integral, the second saddle point approximation performed on the paths brings down an additional amplitude factor canceling out the vanishing $\partial^2 W / \partial \tau^2$. 
Before examining the other two possibilities generating a pole in Eq.~\eqref{eq:FeynmanGarrod}, we stress that we expect to recover a description exclusively in terms of particle trajectories under the two assumptions of eikonal and paraxial. 
The second one in particular, requires that the wave propagates along a main optical axis, with small deviations away from it. 
Therefore, the other two options for the vanishing of the denominator, are excluded from the range of validity of the paraxial assumption: when the velocity vanishes, the particle's trajectory hits a turning point inverting the direction of motion, while when the velocity and the acceleration (or force) are orthogonal, the particle changes direction of propagation. In both cases, the path changes direction considerably, hence invalidating the paraxial assumption.
Intuitively, to transition toward the paraxial approximation we demand that the trajectories included in the path integral always {\it move forward} and we further restrict the trajectories considered for the path integral in Eq.~\eqref{eq:FeynmanGarrod}, already constrained by Eq.~\eqref{Constraint_refractive}, to those that satisfy the paraxial approximation.

\bigskip
\noindent
In the standard literature of WO, the diffraction integral is also used to compute the {\it beyond geometric optics} corrections. 
These are corrections to the $\omega \to \infty$ limit of the amplification factor and are computed by expanding up to fourth order in ${\bf x} - {\bf x}_{cl}$ the phase of the highly oscillating exponential of the diffraction integral. However, from our derivation it is clear that the derivation of Eq.~\eqref{Diffraction_Integral} is approximated in nature, and especially the comparison with the Feynman-Garrod representation in Eq.~\eqref{eq:FeynmanGarrod} questions the validity of the beyond geometric optics expansion. Indeed, in such expansions the proper time $\tau$ and the path ${\bf x}(\tau')$ are treated on unequal footings. One should first make sure that the assumption of $\hat {\cal H} = 0$ is consistent with keeping higher orders in the perturbation series around the classical trajectory.

\section{Perturbation theory}
\label{sec:perturbative_approach}

In the previous Section, we derived the laws of GO and the diffraction integral starting from our main result in Eq.~\eqref{G:path_integral}. 
In this part of the work, we set up the perturbative expansion of the Green function in the parameter $\alpha \omega$, investigating its prediction away from the high-frequency regime.

\subsection{Setting up the the expansion}
Identifying the free action as $S_0 \equiv \int_0^{\tau}d\tau^\prime \, \frac{1}{4}\left(\frac{d{\bf x}}{d\tau^\prime}\right)^2 $, we can rewrite the total Green function of Eq~\eqref{G:path_integral} as
\begin{equation}
    \label{eq:propagator_free_action}
    G_{\omega}({\bf x}_f,{\bf x}_i)=-\frac{i}{\omega}\int_0^{+\infty}d\tau\, e^{i\omega \tau} \int_{{\bf x}(\tau^\prime=0)={\bf x}_i}^{{\bf x}(\tau^\prime=\tau)={\bf x}_f} \mathcal{D}{\bf x}(\tau^\prime)\, e^{i\omega S_0}\exp\left[-i\omega \int_0^\tau d\tau^\prime\,V({\bf x}(\tau^\prime))\right]\,,
\end{equation}
and expand in series the last exponential if the combination of the parameter $\alpha \omega$ is much smaller than one ($\alpha$ is contained in the definition of $V$),
\begin{equation}
    \label{eq:grav_potential_expansion}
    \exp\left[-i\omega \int_0^\tau d\tau^\prime\,V({\bf x}(\tau^\prime))\right] \simeq 1-i \omega \int_0^\tau d\tau^\prime\,  V({\bf x}(\tau^\prime))- \frac{\omega^2}{2}\left(\int_0^\tau d\tau^\prime\, V({\bf x}(\tau^\prime))\right)^2+\dots
\end{equation}
We can use this expression to define the Green function at various perturbation orders
\begin{equation}\label{eq:Propagator_factorization}
     G_\omega({\bf x}_f,{\bf x}_i) = -\frac{i}{\omega}  \int_0^{+\infty}d\tau\, e^{i\omega \tau} \left[ \tilde G^{(0)}_\omega({\bf x}_f,{\bf x}_i,\tau) - i \omega \, \tilde G^{(1)}_\omega({\bf x}_f,{\bf x}_i, \tau) - \frac{\omega^2}{2}\, \tilde G^{(2)}_\omega({\bf x}_f,{\bf x}_i, \tau) + \dots\right]
\end{equation}
with
\begin{align}
    \tilde G^{(0)}_\omega({\bf x}_f,{\bf x}_i, \tau) &\equiv \int_{{\bf x}(\tau'=0) ={\bf x_i}}^{{\bf x}(\tau'=\tau) = {\bf x}_f} \mathcal{D}{\bf x}(\tau^\prime)\, e^{i\omega S_0}\,, \label{eq:PerturbationTheoryZero}\\
    \tilde G^{(1)}_\omega({\bf x}_f,{\bf x}_i, \tau) &\equiv  \int_{{\bf x}(\tau'=0) ={\bf x_i}}^{{\bf x}(\tau'=\tau) = {\bf x}_f} \mathcal{D}{\bf x}(\tau^\prime)\, e^{i\omega S_0}\int_0^{\tau}d\tau^\prime V({\bf x}(\tau^\prime))\,, \label{eq:PerturbationTheoryFirst}\\
    \tilde G^{(2)}_\omega({\bf x}_f,{\bf x}_i, \tau) &\equiv  \int_{{\bf x}(\tau'=0) ={\bf x_i}}^{{\bf x}(\tau'=\tau) = {\bf x}_f}\mathcal{D}{\bf x}(\tau^\prime)\, e^{i\omega S_0}\left[\int_0^\tau d\tau_1 V({\bf x}(\tau_1))\int_0^\tau d\tau_2 V({\bf x}(\tau_2))\right]\label{eq:PerturbationTheorySecond}
\end{align}
Given a specific form of the gravitational potential $V$, one can then use these equations to compute its effects at various order. Note that, because $V = 4 \alpha U$, each order in the expansion is proportional to $(\omega \alpha)^n$. 
In the next two Sections, we will provide further details about the latter equations.

\subsection{The free relativistic particle propagator}

We start by computing the free propagator, namely Eq.~\eqref{eq:PerturbationTheoryZero}. 
In this case, the path integral can be evaluated (see Appendix~\ref{app:FreeProp}) and $G_\omega^{(0)}({\bf x}_f,{\bf x}_i)$ takes the explicit form
\begin{equation}
\label{eq:free_particle_propagator_x_space}
    \tilde G_\omega^{(0)}({\bf x}_f,{\bf x}_i, \tau )= \left(  \frac{ \omega  }{4 i \pi  \tau} \right)^{3/2} \: e^{\frac{i\omega}{4\tau}|{\bf x}_f-{\bf x}_i|^2}\,,
\end{equation}
while the integration over proper time gives
\begin{equation}\label{eq:FreeProp}
    G^{(0)}_\omega ({\bf x}_f,{\bf x}_i) = - \frac{i}{\omega}\int^{+\infty}_0 d \tau\, e^{i \omega \tau} \tilde G_\omega^{(0)}({\bf x}_f,{\bf x}_i, \tau ) =  -\frac{1}{4 \pi}  \frac{e^{i \omega |{\bf x}_f-{\bf x}_i|}}{ |{\bf x}_f-{\bf x}_i|} \,,
\end{equation}
which is the standard form of the Green function of the wave operator in frequency space.
In order to get a more familiar form, one can move to momentum space by performing a 3D Fourier transform
\begin{align}
\label{eq:free_prop_momentum_space}
    G^{(0)}_\omega ({\bf p})&= -\frac{1}{4 \pi}   \int d^3 x \,  \frac{e^{i \omega |{\bf x}_f-{\bf x}_i| -i {\bf p} \cdot ({\bf x}_f - {\bf x}_i)} }{ |{\bf x}_f-{\bf x}_i|} =  \frac{1 }{p^2 - \omega^2}
\end{align}
Drawing an analogy with the propagator of a massive scalar field, we see that $i\omega$ plays the role of the mass of the associated particle \cite{leung2023wave}.

\bigskip
\noindent
If the propagation does not start at $\tau' = 0$ (we will see an example later), the free result is generalized to
\begin{align}
\label{eq:free_particleTauATauB}
    \tilde G_\omega^{(0)}({\bf x}_a,{\bf x}_b, \tau_a- \tau_b )&=  \int_{{\bf x}(\tau'=\tau_b) ={\bf x}_b}^{{\bf x}(\tau'=\tau_a) = {\bf x}_a}\mathcal{D}{\bf x}(\tau^\prime)\, e^{i\omega \int^{\tau_a}_{\tau_b} d \tau' L_0[{\bf x}(\tau'), \dot {\bf x}(\tau'), \tau']} = \nonumber\\
    & =  \left(  \frac{  \omega  }{4 i \pi (\tau_a - \tau_b)} \right)^{3/2} \: e^{\frac{ i \omega}{4 }\frac{|{\bf x}_a-{\bf x}_b|^2}{(\tau_a - \tau_b)}}\,.
\end{align}

\subsection{First order}

We now turn to the first order contribution, namely Eq.~\eqref{eq:PerturbationTheoryFirst}.
In this expression the only information about the specific path ${\bf x}(\tau)$ that the gravitational potential can give is the position of the path at the given proper time $\tau^\prime$. Before and after that moment, the scalar field is evolving freely. With this consideration, we proceed following Feynman's argument in~\cite{Feynman:100771}, and divide each path in two parts (one before $\tau=\tau^\prime$ and one after), and require that, at the given time $\tau_1$, each path goes through a specific point ${\bf x}^\star_1= {\bf x}(\tau_1)$. 
By doing so, and integrating over all possible values of this division point ${\bf x}^\star_1$, one gets the following expression
\begin{equation}
\label{eq:proper_time_second_order}
    \tilde G^{(1)}_\omega({\bf x}_f,{\bf x}_i, \tau)=  \int_0^\tau d\tau_1\,\int_{-\infty}^{+\infty} d{\bf x}^\star_1 \,\tilde G^{(0)}_\omega({\bf x}_f,{\bf x}^\star_1, \tau- \tau_1 )V({\bf x}^\star_1)\tilde G^{(0)}_\omega({\bf x}^\star_1,{\bf x}_i, \tau_1)\,,
\end{equation}
which suggests the interpretation of a particle propagating freely except for when it interacts with $V$. Such interaction can be seen as a ``scattering event'' occurring at a specific point in space. The full probability amplitude is then built from marginalizing over all possible scattering centers.  \\
This representation is widely known in literature. Particularly in the context of cosmology and gravity, it has been used to described the contributions due to self-forces and tail effects \cite{Pfenning:2000zf, Chu:2019vmh, Copi:2022ire,Chu:2011ip}.

\subsection{Higher orders}\label{sec:HigherOrderPTDyson}
\begin{figure}[h!]
    \centering
    \includegraphics[width = \textwidth]{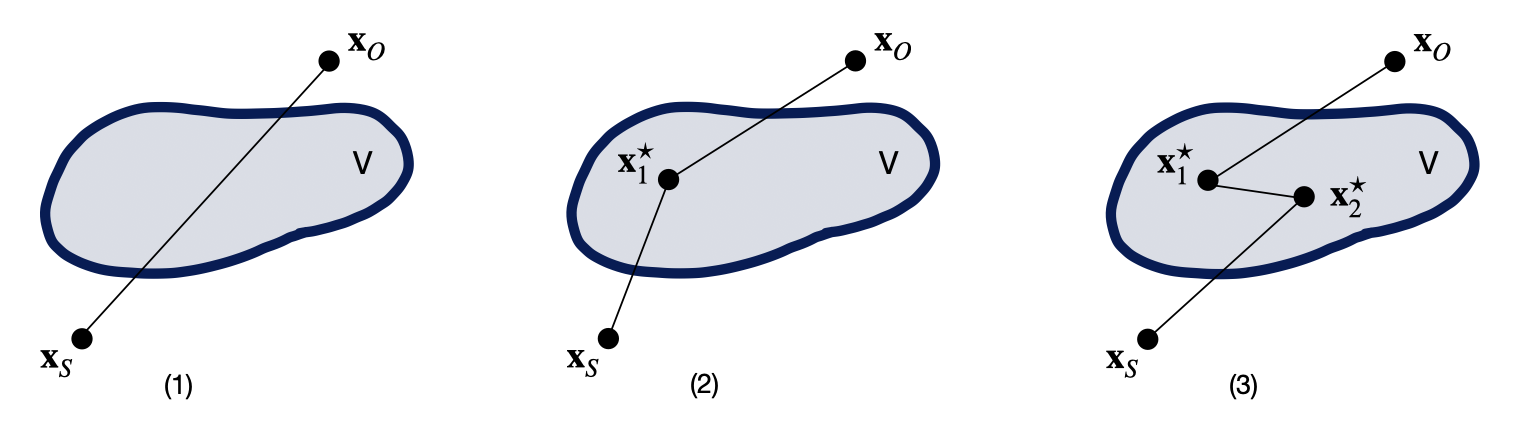}
    \caption{Depiction of the perturbation series we set up, showing three orders in the Born approximation. The solid line represent the free propagator $\tilde G^{(0)}_\omega ({\bf x}_f,{\bf x}_i, \tau)$ connecting various initial and final points. In figure  $(1)$, the particles travels through the region where $V\neq 0$ freely, hence representing the lowest perturbative order.  Adding more scattering points means going to higher orders in perturbation theory. Figures $(2)$ and $(3)$ represent $\tilde G^{(1)}_\omega ({\bf x}_f,{\bf x}_i, \tau)$ and $\tilde G^{(2)}_\omega ({\bf x}_f,{\bf x}_i, \tau)$ respectively, where the wave scatters once and twice with the potential at the points ${\bf x}^\star_1$ and ${\bf x}^\star_2$. The propagation occurs from the source ${\bf x}_i = {\bf x}_S$, to the observer ${\bf x}_f = {\bf x}_O$. In the ideal limit of infinite scattering points, there is no more free propagation inside the potential.}
    \label{fig:Feynman}
\end{figure}

Now we work Eq.~\eqref{eq:PerturbationTheorySecond}.  
As in~\cite{Feynman:100771}, we assume $\tau_1 < \tau_2$ and double the integral. To avoid clutter in the notation, we implement this condition by assuming causal propagation, namely that $\tilde G^{(2)}_\omega ({\bf x}_a,{\bf x}_b, \tau_a - \tau_b ) = 0$ if $\tau_a - \tau_b < 0$.
Then, as done for the first order, we divide the propagation path into three pieces: before $\tau_1$, between $\tau_1$ and $\tau_2$ and after $\tau_2$. This reflects the fact that the particle scatters with the potential at the two proper time instants of $\tau_1$ and $\tau_2$ and propagates freely otherwise. Proceeding as before,  one obtains
\begin{align}
    \tilde G^{(2)}_\omega ({\bf x}_f,{\bf x}_i, \tau) &=  2 \int_0^\tau d\tau_1 \int_0^\tau d\tau_2 \int_{-\infty}^{+\infty} d{\bf x}^\star_1 \int_{-\infty}^{+\infty} d{\bf x}^\star_2 \, \times \nonumber\\
    &~~\times \left[ \tilde G^{(0)}_\omega({\bf x}_f,{\bf x}^\star_1, \tau- \tau_1 )V({\bf x}^\star_1)
    \tilde G^{(0)}_\omega({\bf x}^\star_1,{\bf x}^\star_2, \tau_1-\tau_2) V({\bf x}^\star_2)\tilde G^{(0)}_\omega({\bf x}^\star_2,{\bf x}_i, \tau_2) \right]\,.
\end{align}
Going to higher orders in perturbation theory in the worldline description means adding more and more scattering points. 
Indeed, organizing the perturbation theory in this fashion has the advantage of being straightforwardly iterated to higher orders:  $\tilde G^{(3)}_\omega ({\bf x}_f,{\bf x}_i, \tau)$ will have a similar structure with one more gravitational potential insertion and so on.
This way, at the $n-th$ order in perturbation, the probability of going from ${\bf x}_i$ to ${\bf x}_f$ is built as the sum of the probabilities of doing so with $i$ scattering, where $i \in [0, n]$.
It is important to point out that no assumption has been made on the {\it shape} of the gravitational potential, such as the thin lens approximation. Indeed, here $V({\bf x}(\tau))$ is extended and the integration over all the division points ${\bf x}^\star$ spans all spacetime, also the points where the potential is non-zero. 
Nonetheless, the perturbation theory considers free propagation before and after each scattering event, {\it even if} the propagation is occurring in a region where $V \neq 0$.
This assumption is called {\it Born approximation}~\cite{Feynman:100771}.
The way that the perturbation theory reaches higher levels of accuracy is by considering more and more interactions points between the wave and the potential. 
At higher perturbative orders, the number of scattering points grows, accounting for a less coarse exploration of the potential. This structure is represented in Figure~\ref{fig:Feynman}. In the $n \to \infty$ limit, Born approximation becomes exact as there is no more free propagation inside the gravitational potential.
This suggest that, in this limit, one can resum the perturbation theory and obtain an exact result.
Indeed, by considering infinite scattering points, the expansion series can be organized in such a way that, by factoring out the last scattering point, the Green function takes the form 
\begin{equation}    \label{eq:Gtilde_exact_perturbation_serie}
    \tilde G_\omega ({\bf x}_f,{\bf x}_i, \tau)=\tilde G_\omega^{(0)} ({\bf x}_f,{\bf x}_i, \tau) -i\omega \int_0^\tau d\tau_{LS} \,\tilde G_\omega^{(0)} ({\bf x}_f,{\bf x}_{LS}, \tau- \tau_{LS})\, V({\bf x}_{LS})\,\tilde G_\omega( {\bf x}_{LS},{\bf x}_i,  \tau_{LS})\,.
\end{equation}
This equation states that the full probability amplitude for the associate particle to propagate form the initial ${\bf x}_i$ to the final ${\bf x}_f$ point (the l.h.s), can be obtained as the sum of the free propagation (first term on the r.h.s) and the conditional probability of interacting on the last scattering point (LS) after having propagated with probability amplitude dictated by the full theory. 
Eq.~\eqref{eq:Gtilde_exact_perturbation_serie} is an integral equation for the exact Green function. 
Interpreting  the gravitational potential $V({\bf x})$ as the proper self-energy, then
Eq.~\eqref{eq:Gtilde_exact_perturbation_serie} becomes equivalent to  {\it Dyson equation}, found in quantum field theory~\cite{Peskin:1995ev} and in non-relativistic many body theory~\cite{Fetter}. 
The interpretation of the gravitational potential as self-energy, dressing the free propagator, is also clear from the Hamiltonian description laid out in Section~\ref{sec:Hamilton}.

\subsection{Perturbation theory with the Generating Functional}

A key feature of the path integral description, is its organization in terms of the generating functional~\cite{schwinger,Schwartz:2014sze, Srednicki:2007qs}. 
By adding a source to the phase weighting the paths, one can obtain expectation values of operators by taking functional derivatives of the generating functional with respect to it. 
In the context of the worldline representation, the generating functional is defined as 
\begin{equation}
\label{eq:Generating_functional}
    Z[{\bf J}, \tau]= \int_{{\bf x}(\tau' = 0) ={\bf x}_i}^{{\bf x}(\tau' = \tau) = {\bf x}_f} \mathcal{D}{\bf x}(\tau') \exp \left[i\omega \int_0^\tau d\tau^\prime \left(L+{\bf J}(\tau^\prime) \cdot {\bf x}(\tau^\prime)\right)\right]\,,
\end{equation}
where $L$ is the associated Lagrangian, given by Eq. \eqref{eq:lagrangian} and ${\bf J}(\tau')$ is the current acting as the source of ${\bf x}$.
Note that, since we are working at the particle level through the worldline approach, operators are functions of ${\bf x}(\tau)$, rather than fields. For instance, in Eq.~\eqref{eq:PerturbationTheorySecond}, we have two gravitational potentials evaluated in ${\bf x}(\tau_1)$ and ${\bf x}(\tau_2)$. 
Expectation values of operators can be obtained by taking functional derivatives with respect to the current ${\bf J}$ and then setting it to zero, namely
\begin{equation}
  \frac{\delta^n  Z[{\bf J}, \tau]}{ \delta {\bf J}(\tau_1) \dots \delta {\bf J}(\tau_n)} \, \Bigg|_{{\bf J = 0}} = \int_{{\bf x}(\tau'=0) ={\bf x}_i}^{{\bf x}(\tau'=\tau) = {\bf x}_f}\mathcal{D}{\bf x}(\tau^\prime)\, e^{i\omega S} \, {\bf x}(\tau_1) \, \dots {\bf x}(\tau_n) \,,
\end{equation}
since the action of the functional derivative is analogous to its counterpart in quantum field theory~\cite{Peskin:1995ev}, namely $\delta J_i(\tau_1) / \delta J_j(\tau_2) = \delta^j_i \, \delta(\tau_1-\tau_2)$, thus 
\begin{equation}
    \frac{1}{i \omega}\frac{\delta}{\delta {\bf J}(\tau_1)}\exp \left[i\omega \int_0^\tau d\tau^\prime {\bf J}(\tau^\prime){\bf x}(\tau^\prime)\right]= {\bf x}(\tau_1)\exp \left[i\omega \int_0^\tau d\tau^\prime {\bf J}(\tau^\prime) \cdot {\bf x}(\tau^\prime)\right]\,.
\end{equation}
As usual, one can produce a perturbative expansion also through the generating functional by writing the latter as
\begin{align}
\label{eq:Generating_functional_perturbative}
    Z[{\bf J}, \tau]&= \exp \left[i\omega \int_0^\tau d\tau^\prime V\left(\frac{1}{i \omega}\frac{\delta}{\delta {\bf J}}\right)(\tau^\prime)\right] Z_0[{\bf J}, \tau]\,,
\end{align}
where $Z_0[{\bf J}, \tau]$ is the generating functional of the free theory,
\begin{align}
    \label{eq:Generating_free_theory}
    Z_0[{\bf J}, \tau]&\equiv \int_{{\bf x}(\tau' = 0) ={\bf x}_i}^{{\bf x}(\tau' = \tau) = {\bf x}_f} \mathcal{D}{\bf x}(\tau') \exp \left[i\omega \int_0^\tau d\tau^\prime \left(\frac{\dot{\bf x}^2}{4}+{\bf J}(\tau^\prime) \cdot{\bf x}(\tau^\prime)\right)\right]\,.
\end{align}
In Eq.~\eqref{eq:Generating_functional_perturbative} the operator $V$ evaluated in the derivative with respect to the current, is to be intended through its Taylor expansion.
The free generating functional is quadratic in the velocity, hence it can be computed explicitly
\begin{align}\label{eq:completing the square}
    Z_0[{\bf J}, \tau] &= \exp \left[i\omega \int_0^\tau d\tau^\prime \left(-\frac{1}{4}{\bf x}\frac{d^2}{d\tau^2}{\bf x}+{\bf J}(\tau^\prime){\bf x}(\tau^\prime)\right)\right] =\nonumber\\
    &=\exp \left[i\omega \int_0^\tau d\tau^\prime \mathcal{A} \left(-\frac{1}{2}{\bf x}-{\bf J}(\tau^\prime)\mathcal{A}^{-1}\right)^2\right]\exp \left[i\omega \int_0^\tau d\tau^\prime d\tau^{\prime\prime} {\bf J}(\tau^\prime)\Delta(\tau^\prime-\tau^{\prime\prime}){\bf J}(\tau^{\prime\prime})\right]\,,
\end{align}
where $\mathcal{A} \equiv \frac{d^2}{d\tau^2}$  and $\Delta(\tau^\prime-\tau^{\prime\prime})$ can be interpreted as a {\it proper time propagator},
describing the probability to go from $\tau$ to $\tau'$ and obeying
\begin{equation}
    \label{propagator in tau}
    \frac{d^2}{d\tau^2}\Delta(\tau^\prime-\tau^{\prime\prime})=\delta(\tau^\prime-\tau^{\prime\prime})\,.
\end{equation}
Note that the first factor in Eq.~\eqref{eq:completing the square} gives an overall normalization constant, which can be reabsorbed in the overall one.

\section{Applications}\label{sec:Applications}
In this Section we provide two applications of our formalism. First we show how it can be implemented also for massive scalar fields. 
Then, we provide an explicit example of the first order correction to the propagator, i.e. we compute Eq.~\eqref{eq:proper_time_second_order}, for a specific choice of gravitational potential.

\subsection{Coulomb potential: an explicit example}\label{sec:Yukawa}
\begin{figure}[t!]
    \centering
    \includegraphics[width = \textwidth]{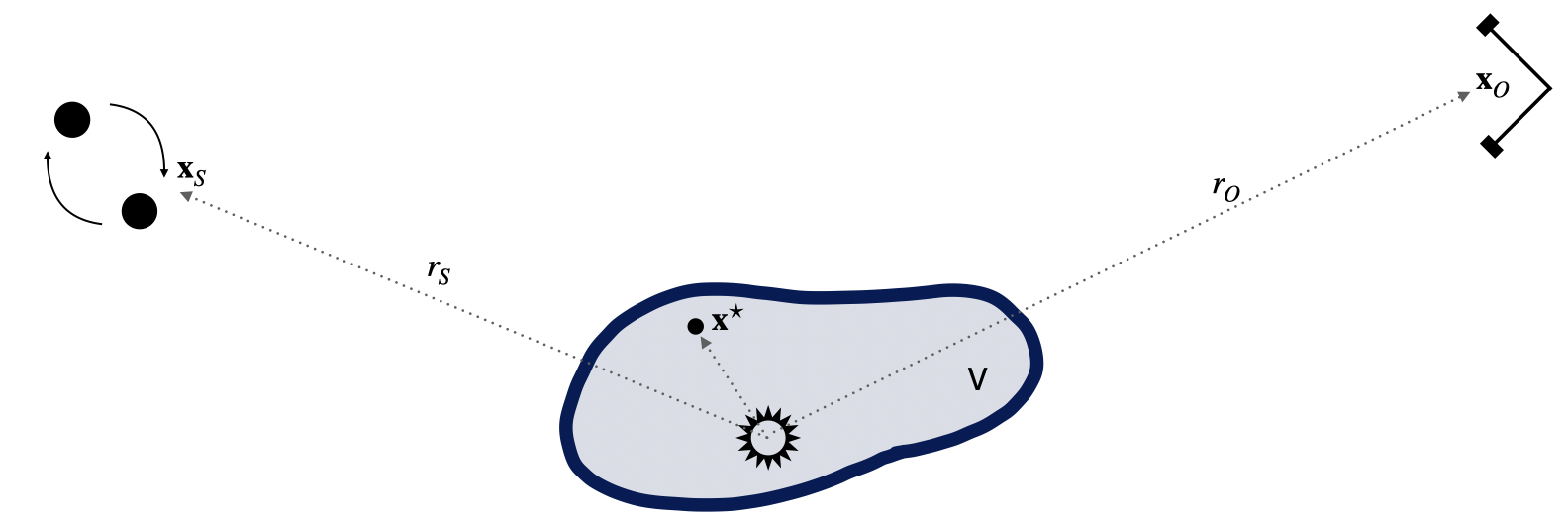}
    \caption{Lensing situation considered. The coordinate system is now centered on the lens and the distances to the source and observer are large.}
   \label{fig:LensigCoulomb}
\end{figure}

In this section, we obtain an explicit result for the case of a spherically symmetric potential, describing the standard scenario of a gravitational potential generated by a compact object. In particular, we shall compute $\tilde G^{(1)}_\omega({\bf x}_f,{\bf x}_i, \tau)$ of Eq.~\eqref{eq:proper_time_second_order} for a Coulomb-like potential, i.e. $U\sim 1/r$, and the total first order Green function in frequency space, $ G^{(1)}_\omega({\bf x}_f,{\bf x}_i)$, by integrating over the proper time. 
To do so, we will first consider a Yukawa potential,
\begin{equation}\label{eq:Yukawa potential}
    U = -  \frac{M G }{r} e^{- r/a}\,,
\end{equation}
where $a$ sets the scale of the range of the interaction, and then take the $a \to \infty$ limit.
Introducing this length-scale, allows us to regularize the divergent integral in Eq.~\eqref{eq:proper_time_second_order}, and perform it. 
We also recall that, in the unit where $c=1$, then $M G$ has dimension of a length and it coincides with the Schwarzschild radius of the lens, and the gravitational potential is dimensionless. 
In order to have a more clear picture in mind, let us identify the initial propagation point with the position of the source emitting the GW (i.e. ${\bf x}_i = {\bf x}_S $) and the final point with the position of the observer (i.e. ${\bf x}_f ={\bf x}_O $).
It is convenient to center the coordinate frame on the lens, and consider that $r_S$ and $r_{O}$ are much larger that the distance to the scattering point $r_{\star}$.
Namely, we consider that the points where the waves are produced and observed are distant from the point where the interaction occurs. This is a good approximation when $V({\bf x})$ has a compact support, or decays fast enough at large distances, hence allowing us to stop at first order in perturbation theory. 
In particular, we consider (see Figure~\ref{fig:LensigCoulomb}) 
\begin{equation}
    {\bf x}_S = \{r_S, - \boldsymbol{\theta}_S \}\,, \qquad  {\bf x}_O = \{r_O, \boldsymbol{\theta}_O \}\,, \qquad {\bf x}^\star = \{r^\star, \boldsymbol{\theta}^\star \}\,.
\end{equation}
The integral over time of Eq.~\eqref{eq:proper_time_second_order} can be performed 
\begin{align}\label{eq:IntegralGsquare}
    \int^\tau_0 \, d \tau_1 \, &\tilde G^{(0)}_\omega({\bf x}_O,{\bf x}^\star, \tau-\tau_1 ) V({\bf x}^\star) \tilde G^{(0)}_\omega({\bf x}^\star,{\bf x}_S, \tau_1) = \nonumber \\
    & = \left(\frac{\omega}{4 i \pi } \right)^{5/2} \, \tau^{-3/2} \, \left( \frac{1}{|{\bf x}_O - {\bf x}^\star|} + \frac{1}{|{\bf x}_S - {\bf x}^\star|}\right) \: e^{\frac{i \omega}{4 \tau} \left( |{\bf x}_O - {\bf x}^\star| + |{\bf x}_S - {\bf x}^\star|\right)^2 } V({\bf x}^\star)\,,
\end{align}
where we have used that $V({\bf x}^\star)$ does not depend on $\tau'$ at this stage (see Appendix~\ref{app:Coulomb} for the explicit computation). To proceed with the spatial integration we consider that the source and observer are located far away compared to the lens. Imposing $r_S \gg r^\star$ and $r_O \gg r^\star$ gives
\begin{align}\label{eq:SOfarfromL}
   |{\bf x}^\star-{\bf x}_S| \simeq  r_S \left(1 + \frac{r^\star}{r_S} \,\boldsymbol{\theta}_S\cdot \boldsymbol{\theta}^\star \right)\,, \qquad
    |{\bf x}^\star-{\bf x}_O| \simeq r_O \left(1 - \frac{r^\star}{r_O}\, \boldsymbol{\theta}_O\cdot \boldsymbol{\theta}^\star \right)\,.
\end{align}
which we keep at lowest order in the first parenthesis of Eq.~\eqref{eq:IntegralGsquare}, and at first order in the exponential since this is quite sensitive to small relative changes in the phase. 
As prescribed in Eq.~\eqref{eq:proper_time_second_order}, we compute the spatial integral in the case of the Yukawa potential  in Eq.~\eqref{eq:Yukawa potential}. Defining the vector ${\bf R} \equiv (r_S + r_O)(\boldsymbol{\theta}_S - \boldsymbol{\theta}_O)$, with modulus $R \equiv \sqrt{|{\bf R}|^2}$ the first order worldline propagator takes the form (see Appendix~\ref{app:Coulomb} for the explicit steps) 
\begin{align}\label{eq:FirstGreenProperTimeYukawa}
     \tilde G^{(1)}_\omega({\bf x}_O,{\bf x}_S, \tau)&=  \int d^3 {\bf x}^\star \int^\tau_0 \, d \tau' \, \tilde G^{(0)}_\omega({\bf x}_O,{\bf x}^\star, \tau-\tau' ) V({\bf x}^\star) \tilde G^{(0)}_\omega({\bf x}^\star,{\bf x}_S, \tau') \nonumber \\
     &\approx 16 \pi \alpha MG \left( \frac{\omega}{4 i \pi}\right)^{5/2}  \left( \frac{1}{r_O} + \frac{1}{r_S}\right)  \times   \frac{\tau^{-3/2}\, e^{\frac{i \omega}{4 \tau} (r_S + r_O)^2} }{\left( \frac{i \omega R}{2 \tau}\right)^2 - \frac{1}{a^2}} \,.
\end{align}
We see that the scale of the interaction $a$, can be removed at this point by taking $a\to \infty$, as the divergent integral has now been performed. We take this limit and produce the result for the full propagator, by integrating out the proper time (see Appendix~\ref{app:Coulomb} for the explicit steps)
\begin{align}\label{eq:FirstGreenYukawa}
     G^{(1)}_\omega({\bf x}_O,{\bf x}_S) &\equiv -\frac{i}{\omega} \, \, \int^{+\infty}_0 d \tau \: e^{i \omega \tau} \: \tilde G^{(1)}_\omega({\bf x},{\bf x}_0, \tau) = \nonumber\\
     &=\frac{ \alpha MG}{\pi \omega^2} \left( \frac{1}{r_O} + \frac{1}{r_S}\right)  \frac{e^{i \omega (r_S + r_O)} (1 - i \omega (r_S + r_O))}{R^2}\,.
\end{align}
Hence we have arrived to the first order Born approximation of the Green function describing the dynamics of the system, which considers that the wave propagates freely except when the one interaction takes place. 
We can compose the full Green function by summing the zero-th and first order results for the Coulomb potential 
\begin{align}\label{eq:GreenTotCoulomb}
    G_\omega ({\bf x}_O,{\bf x}_S) &=  G^{(0)}_\omega ({\bf x}_O,{\bf x}_S)  - i \omega  G^{(1)}_\omega ({\bf x}_O,{\bf x}_S) \approx \nonumber \\
    &\approx-\frac{1}{4\pi} \frac{e^{i \omega (r_O + r_S)}}{r_O + r_S}  \left[1  + 4 i \alpha (\omega M G) \, \frac{1 - i \omega(r_S +r_O)}{r_S r_O |\boldsymbol{\theta}_S -  \boldsymbol{\theta}_O|^2 \omega^2}\right]
\end{align}
where we have used Eq.~\eqref{eq:FreeProp} and in the last line we have written the parenthesis in a way that clearly shows that the first order contribution is dimensionless ($MG$ is the Schwarzschild radius of the lens).
We point out some characteristic of Eq.~\eqref{eq:GreenTotCoulomb}. First, the divergence in the forward limit $\boldsymbol{\theta}_S \approx  \boldsymbol{\theta}_O$ is typical of an interaction mediated by a long range force, inducing an interaction between the wave and the lens also at very long distances, and it is mitigated by the smallness of the interaction (the coupling constant is $G$). 
Another remarkable feature of Eq.~\eqref{eq:GreenTotCoulomb} regards the static limit,  which seems not to be possible for the $1/\omega$ factor. However, in our treatment we had to assume $\omega \neq 0$ in order to introduce the proper time propagator with the correct $- i / \omega$ normalization. Because of this reason, taking the static limit is subtle but one can do so by considering $\omega \to 0$ while keeping $ \omega r_S , \omega r_O $ finite. This prescription is necessary to remain within the realm of our assumption, and it leads to the correct results as well. 
Another possibility, is to take the limit $\omega M G \to 0$ as proxy for the static limit. In this case, the wavelength of the wave ($\lambda \sim 1/ \omega$) becomes much larger than the Schwarzschild radius of the lens and the interaction is suppressed as expected.
Indeed, we see that the interaction (encoded in $G^{(1)}_\omega ({\bf x}_O,{\bf x}_S)$) vanishes either if $\alpha = 0$, i.e. when there is no lens, or in the $M G / \lambda \to 0$  limit which corresponds to the fully diffractive regime.
We also point out that any frequency dependent modification to the Green function appears in combination $i \omega$ or $\omega^2$ compatibly with the fact that the Green function in time domain is real, implying $G_\omega = G_{-\omega}^*$.

Finally, we use our result to connect our theoretical framework to an observable, namely the distance to the source. This can be obtained from the waveform through an inverse proportionality relation in GO regime~\cite{Isaacson1,Isaacson2}. The presence of the lens along the propagation will introduce modifications in the inferred distance, which are also direction dependent~\cite{Bertacca:2017vod}. 
We consider a  source ${\cal S}$ localized at ${\bf x}_s$, so that the scalar field at the observer position takes the form $\tilde \Psi_\omega ({\bf x}_O) = {\cal A}_{in} \, G_\omega ({\bf x}_O, {\bf x}_S)$, where ${\cal A}_{in}$ is the initial amplitude. 
From the form of the free propagator, we recognize in $r_S + r_O$ the distance from the source to the observer, since we set the center of the coordinate system on the lens. We call this distance $\bar d_{SO} \equiv r_S + r_O$, which is the unperturbed distance in the sense that it is the correct one in the absence of the lens, and it does not depend on the direction of arrival of the wave.
We can then use the fact $\alpha \omega \ll 1$ to rewrite the propagator in Eq.~\eqref{eq:GreenTotCoulomb} as 
\begin{align}\label{eq:PsiCoulombFirstOrder}
    \tilde \Psi_\omega ({\bf x}_O)  \approx -\frac{{\cal A}_{in}  }{4 \pi}\,  \frac{e^{i \omega (r_S + r_O)}}{\bar d_{SO} \left[1  -4 i \alpha (\omega M G) \, \frac{1 - i \omega (r_S + r_O)}{r_S r_O |\boldsymbol{\theta}_S -  \boldsymbol{\theta}_O|^2 \omega^2}\right]} = -\frac{{\cal A}_{in}  }{4 \pi}\,  \frac{e^{i \omega (r_S + r_O)}}{d_{SO} } \,, 
\end{align}
where in the last step we identified the denominator as the corrected distance in presence of the lens, $d_{SO}$.  While the unperturbed distance depends only on the radial coordinates, the presence of the lens introduces a dependence of $d_{SO}$ also on the deviation angle. 
We can compute the source-observer distance fluctuation as
\begin{equation}\label{eq:DistanceFluctuations}
    \frac{d_{SO} - \bar d_{SO}}{\bar d_{SO}} \equiv \frac{\Delta d_{SO}}{\bar d_{SO}} = -4 i \alpha (\omega M G) \, \frac{1 - i \omega (r_S + r_O)}{r_S r_O |\boldsymbol{\theta}_S -  \boldsymbol{\theta}_O|^2 \omega^2}\,.
\end{equation}
To facilitate the comparison with other works in literature, we also report the expression of the amplification factor which can be understood from Eq.~\eqref{eq:PsiCoulombFirstOrder}. This is given by 
\begin{equation}\label{eq:AmplFactorCoulomb}
    F_\omega( {\bf x}_O) = \frac{\tilde \Psi_\omega ({\bf x}_O)}{\tilde{\Psi}^{NL}_\omega ({\bf x}_O)} = 1 + 4 i \alpha (\omega M G) \, \frac{1 - i \omega (r_S + r_O)}{r_S r_O |\boldsymbol{\theta}_S -  \boldsymbol{\theta}_O|^2 \omega^2}
\end{equation}
where the unlensed solutions can be found by setting $M = 0$ in Eq.~\eqref{eq:PsiCoulombFirstOrder}. One can compare this expression for the amplification factor to its analogous result obtained from the diffraction integral (see e.g.~\cite{Tambalo:2022plm} Eq.~$22$) . 
Both amplification factors show that deviations from the unlensed solution are proportional to $\omega M G$, indicating the vanishing of the effects in the fully diffractive regime, and their structure in the complex frequency plane is such that the real time counterpart is real.   Both expressions show a divergence in the $\omega \to 0$ limit, in Eq.~\eqref{eq:AmplFactorCoulomb} with a $1 / \omega$ falloff, and in~\cite{Tambalo:2022plm} through a logarithmic term. However, the radial dependencies are different, especially concerning the divergence in the forward limit, due to the different approaches and approximations employed.

\subsection{Massive scalar field}\label{sec:MassiveScalarField}

We now consider a massive scalar field as prototype of massive gravity theory. This case was already considered in~\cite{Morita:2019sau} to investigate the possibility of constraining the field's mass through lensing observations of combined GW and electromagnetic signals. 
In a standard lensing scenario, electromagnetic waves are usually in the GO regime, while GWs can have longer wavelengths and display WO effects. In these situations, the two lensing patterns can be used to break degeneracies and constraint both the properties of the lens and of the two messengers.
Also~\cite{Morita:2019sau} relies on the diffraction integral for its prediction, hence here we generalize their description within our framework, which we can easily re-propose with suitable modifications.

We start from a KG equation for a massive scalar field $ (\bar \Box - m^2 )\Psi = 0 $ in the curved background. The Green function Eq.~\eqref{Green_equation} then becomes
\begin{equation}\label{eq:GreenMassive}
    \left[\nabla^2 +\omega^2 \left( 1 - \frac{m^2}{\omega^2} \right) \left[ 1 -  4 \alpha U \right] \right]G_\omega({\bf x}_f,{\bf x}_i)=\delta^{(3)}\left({\bf x}_f-{\bf x}_i\right)\,.
\end{equation}
We can make the frequency redefinition
\begin{equation}\label{eq:FrequencyMassive}
    \omega_{m} \equiv \omega \, \sqrt{ 1 - \frac{m^2}{\omega^2}}\,,
\end{equation}
and obtain an equation which is exactly equal to Eq.~\eqref{Green_equation} upon replacing $\omega$ with $\omega_{m}$. Consequently, the exact solution of Eq.~\eqref{eq:GreenMassive} is given by  
\begin{equation}
\label{G:path_integral_massive}
    G_{\omega_{m}}({\bf x}_f, {\bf x}_i)=-\frac{i}{\omega_m}\int_0^\infty \, d\tau\, e^{i\omega_{m} \tau}  \int_{{\bf x}(\tau^\prime=0)={\bf x}_i}^{{\bf x}(\tau^\prime=\tau)={\bf x}_f} \mathcal{D}{\bf x}(\tau^\prime) e^{i\omega_{m} S}\,,
\end{equation}
and the discussion of the previous sections can be reproduced straightforwardly but, in this case, the optical regime of the wave is regulated by the parameter $\omega_{m}$ instead of $\omega$.
When the frequency of the wave is much larger than the field's mass, i.e. $\omega^2 \gg m^2$, then $\omega_{m} \to \omega$. 
This regime corresponds to Fourier modes whose associated length-scale ($\sim 1/\omega$) is much smaller than the Compton wavelength of the field, hence they are not influenced by the presence of the new scale. Vice versa, when $\omega^2 \ll m^2$, then the frequency $\omega_m$ becomes purely imaginary and the dynamic is suppressed.

\section{Lensing by a Black Hole: adding the polarization}
\label{sec:spin2 for spheric lenses}

Having set the formalism for a scalar field, in this section we generalize our results for perturbations of higher spin. 
This task would require to solve Maxwell's equations for fields of spin $s=1$, or the linearized Einstein's equations for $s=2$, over the background described in Eq.~\eqref{eq:background_metric}. This task is highly complicated and not analytically solved for generic gravitational potentials yet. 
Nonetheless, for specific forms of $V$, the solutions of the mentioned equations are known in literature. One example are Kerr and Schwarzschild BHs, which we consider in this Section.
To be consistent with the literature, in this Section we consider units where $c = G = 1$. To restore the dependence upon Newton's gravitational constant, it is sufficient to perform $ M \to M G$.

\subsection{Preliminary about Black Hole perturbation theory}

Studying the equations of motion of test fields of different spin is the heart of the literature concerning BH perturbation theory. 
The field has started with the pioneering works of Regge and Wheeler~\cite{Regge:1957td} and Zerilli~\cite{Zerilli:1970se} for Schwarzschild lenses, and it was then expanded to Kerr BHs by Teukolsky in \cite{Teukolsky:1973ha}, providing a master equations for perturbations of different spins over such background.  The key feature of Teukolsky equation is that, rather than being written in terms of the field variables, it is an equation for a specific Newman-Penrose (NP) scalar~\cite{Newman:1961qr}. Such variables are related to derivatives of the test fields and, at linear order in perturbation theory, they are gauge invariant~\cite{Chandrasekhar:1985kt}.
For instance, for $s=2$, Teukolsky equation is an equation for $\psi = \{ \psi_0 \, , \rho^{-4} \psi_4\}$ where $\rho = -1/ (r - i a \cos \theta)$,  $a$ being the Kerr parameter of Eq. \eqref{eq:KerrMetric}, and 
\begin{equation}
    \psi_0 = - \delta \left( C_{\mu  \alpha \nu \beta } \, l^\mu m^\alpha l^\nu m^\beta \right) \,, \qquad  \psi_4 = - \delta \left( C_{\mu  \alpha \nu \beta } \, n^\mu \bar m^\alpha n^\nu \bar m^\beta \right) \,, \qquad  
\end{equation}
where $  C_{\mu  \alpha \nu \beta }$ is the Weyl tensor and $\{ l^\mu, n^\mu, m^\mu, \bar m^\mu \}$ are the elements of the NP null tetrad. As an example, the two  scalars $\psi_0$ and $\psi_4$ on a Minkowski background, are related to gravitational waves as $ \{\psi_0 \,, \psi_4 \} \propto \{ \ddot  h_{+ } + i \ddot h_{\times } \,, \ddot   h_{+ } - i \ddot h_{\times } \}$~\cite{Maggiore:2018sht}, namely the second time derivative of the two GW polarizations. For a vector field, the NP scalars composing the doublet $\psi$ are related to NP components of the Maxwell tensor, while for $s=0$, $\psi$ is proportional to the scalar field itself. 
In this section, therefore, we borrow the results of BH perturbation theory studies, and generalize the formalism presented in the previous Sections, for perturbations of spin $|s| = 0, 1/2 , 1, 2$. 
Hence, here we consider as background metric the one of a Kerr BH
\begin{equation}\label{eq:KerrMetric}
    d \bar s^2 = \left( 1 - \frac{2 M r}{\Sigma} \right) d t^2 + \frac{4 M a r \sin^2 \theta}{\Sigma} d t \, d \varphi - \frac{\Sigma}{\Delta} d \theta^2 - \sin^2 \theta \left( r^2 +  a^2 + \frac{2 M a^2 r \sin^2 \theta}{\Sigma}\right) d \varphi^2\,,
\end{equation}
with 
\begin{equation}
    \Sigma = r^2 + a^2 \cos^2 \theta \,, \qquad \Delta = r^2 - 2 M r + a^2\,,
\end{equation}
instead of the one in Eq. \eqref{eq:background_metric}. 
Teukolsky master equation then reads~\cite{Teukolsky:1973ha,Chandrasekhar:1985kt} 
\begin{align}\label{eq:Teukolsky}
    \Bigg \{ &\left[ \frac{(r^2 + a^2)^2}{\Delta} - a^2 \sin^2 \theta\right] \frac{\partial^2}{\partial t^2} + \frac{4 M a r}{\Delta} \frac{\partial^2 }{\partial t \partial \varphi} +  \left[ \frac{a^2}{\Delta} - \frac{1}{\sin^2 \theta}\right] \frac{\partial^2}{\partial \varphi^2}  - \frac{1}{\sin \theta} \frac{\partial}{\partial \theta} \left( \sin \theta \frac{\partial}{\partial \theta}\right)- \nonumber \\
    & -  2 s \left[ \frac{a(r-M)}{\Delta} + \frac{i \cos \theta}{\sin^2 \theta}\right] \frac{\partial}{\partial \varphi} - 2 s \left[ \frac{M (r^2 - a^2)}{\Delta} - r - i a \cos \theta \right]  \frac{\partial}{\partial t} + \nonumber \\
    &+(s^2 \cot^2 \theta - s) - \Delta^{-s} \frac{\partial}{\partial r} \left( \Delta^{s+1} \frac{\partial}{\partial r}\right) \Bigg \} \, \psi (t,r,\theta,\varphi) = 0\,,
\end{align}
where $\psi$ is the NP scalar, as explained.
It is straightforward to show that the latter equation  reduces to  Eq.~\eqref{Helmholtz} for a spherically symmetric ($a = 0$), static lens in Newtonian regime ($r \gg 2 M$), by taking the suitable limit. 
Indeed, the metric in Eq. \eqref{eq:background_metric} is included in the Kerr family if one restricts the study to radial potentials of the form, i.e. $ U (r, \theta, \varphi) = - M/ r$. 
However, we emphasize  that Eq.~\eqref{eq:SchrodingerOneD} is exact while Helmholtz Eq.~\eqref{Helmholtz} for the scalar problem was found by linearizing to first order in $\alpha$, namely the strength of the gravitational potential.

\subsection{Proper time solution of Teukolsky equation}

The goal of this Section is to show how to recast Eq.~\eqref{eq:Teukolsky} into a form similar to Eq.~\eqref{Helmholtz}, so that one can re-apply all the previous arguments in the case of higher spin fields. 
Following the standard procedure, we exploit the symmetries of the Kerr metric and decompose the scalar variable $\psi$ as
\begin{equation}
    \psi = e^{- i \omega t} \, e^{i m \varphi} \, S(\theta) \, R(r)\,,
\end{equation}
so that Teukolsky equation decouples into two, one for the radial and one for the angular angular variable
\begin{align}
    &\Delta^{-s} \, \frac{d}{d r} \, \left( \Delta^{s+1} \, \frac{d R}{d r} \right) +  V_R(r) R = 0\,, \label{eq:RadialTeukolsky}\\ 
    &\frac{1}{\sin \theta} \, \frac{d}{ d \theta } \left( \sin \theta \, \frac{d S}{d \theta}\right) + V_S (\theta)S =0 \,, \label{eq:AngularTeukolsky}
\end{align}
with the potentials given by 
\begin{align}
    V_R (r) &\equiv \left( \frac{K^2 - 2 i s (r - M) K}{\Delta} + 4 i s \omega r - a^2 \omega^2 + 2 a m \omega -A \right)\,, \\
    V_S (\theta) &\equiv \left( a^2 \omega^2 \cos^2 \theta - \frac{m^2}{\sin^2 \theta} - 2 a \omega s \cos \theta - \frac{2 m s \cos \theta}{\sin^2 \theta} - s^2 \cot^2 \theta + s + A \right) \,,
\end{align}
where $ K \equiv (r^2 + a^2 ) \omega - a m$ and $A$ is the separation constant~\cite{Press:1973zz}, found as the eigenvalue problem relative to Eq.~\eqref{eq:AngularTeukolsky}. 
For non-rotating BH ($ a = 0$),  the angular Teukolsky equation admits as solution spin-weighted spherical harmonics, ${}_s Y_{\ell m}  = {}_s S_{\ell m} (\theta) \, e^{i m \varphi} $, and $A = \ell (\ell + 1) - s (s+1)$. For rotating BH, the solution of the angular equation are ``spin-weighted spheroidal harmonics'', and the separation constant $A$ are known perturbatively for $a \omega \ll 1$, and are given by $A = \ell (\ell + 1) - s(s+1) - 2 a \omega s^2 m /(\ell^2 + \ell) + {\cal O} (a^2 \omega^2)$~\cite{Press:1973zz}. 
Because the solutions of the angular problem are known, we treat the radial Eq.~\eqref{eq:RadialTeukolsky}. This can be turned into a 1 dimensional Helmholtz equation, similar to Eq.~\eqref{Helmholtz}, by making the variable redefinition
\begin{equation}
    R(r) =\Delta^{-\frac{s+1}{2}} \,  \tilde R(r) \,.
\end{equation}
In terms of $\tilde R$, Eq.~\eqref{eq:RadialTeukolsky} becomes
\begin{equation}\label{eq:SchrodingerOneD}
    \frac{d^2 \tilde R}{d r^2} + \omega^2 \left[ 1 - 4 \tilde U^{s}_{\ell m} (\omega, r) \right] \tilde R = 0\,,
\end{equation}
with 
\begin{equation}\label{eq:4OmegaUTilde}
    4 \, \omega^2 \tilde U^{s}_{\ell m} (\omega, r) = \omega^2 +\frac{(1+s) (a^2 - M^2 + (M-r)^2 s) }{\Delta^2}   - \frac{V_R(r)}{\Delta}\,.
\end{equation}
Note that to write the equation above we collected out a factor of $\omega^2$, hence we have implicitly assumed $\omega \neq 0$. 
From the expression in Eq.~\eqref{eq:4OmegaUTilde} we can already point out one interesting new feature of lensing by BH, compared to a Newtonian lens: differently from the potential $V(r)$, $\tilde U^{s}_{\ell m} (\omega, r)$ depends also on the frequency $\omega$, and not simply on the radial direction. This important difference is introduced both by near-horizon effects, or by the rotation of the BH or by the spin of the perturbation, phenomena that were neglected in the previous discussion but were already predicted in literature~\cite{Oancea:2023hgu, Oancea:2022szu, Andersson:2020gsj}.
Writing the Teukolsky equation as in Eq.~\eqref{eq:SchrodingerOneD}, allows us to apply immediately all the tools developed in the previous Sections also in this context, and find a solution of the BH perturbation equation in the worldline representation.
In particular, the Green function of Eq.~\eqref{eq:SchrodingerOneD} will be written as 
\begin{equation}\label{eq:GProperTimeBH}
   G_\omega(r_f, r_i)=-\frac{i}{\omega}\int_0^\infty d\tau \: e^{i \omega \tau }\int_{r(\tau^\prime=0)=r_i}^{r(\tau^\prime=\tau)=r_f} \mathcal{D}r(\tau^\prime) \: e^{i\omega  \int_0^\tau d\tau^\prime \left[\frac{1}{4}\left(\frac{\text d r}{\text d \tau^\prime}\right)^2-V^{s}_{\ell m}(\omega, r)\right]}\,,
\end{equation}
with potential given by
\begin{align}\label{eq:PotentialBH}
    V^{s}_{\ell m}(\omega, r) = 1 - \left( \frac{-(1+s) (a^2 - M^2 + (M-r)^2 s) }{\omega^2 \Delta^2} + \frac{V_R(r)}{\omega^2  \Delta} \right)\,.
\end{align}
With these two equations, we have successfully generalized the standard wave optics treatment for scalar waves to a field with generic spin $s$, and without requiring the paraxial approximation, nor the Newtonian expansion. Eq.~\eqref{eq:GProperTimeBH} is valid for any frequency $\omega$ and field spin $|s| = 0, 1/2, 1, 2$, when the lens is a BH of the Kerr family also close to the horizon.

\subsection{Weak field limit and spin effects}
To compare the results of this Section with the previous ones, we take the weak field limit ($r \gg 2 M$) for non-rotating ($a=0$) BH. This will allow us to understand the role of the spin, which was not accounted for in the standard description of the diffraction integral. 
In the Newtonian limit $\Delta \approx r^2$ and it can be showed that
\begin{align}\label{eq:PotentialSpinNewton}
      4  \tilde U^{s}_{\ell m} (\omega, r) \, \approx \, - 4 \frac{M}{r} +\frac{\ell(\ell + 1) + s(s+1)}{ \omega^2 r^2}   - \frac{ 2 i s}{ \omega r}\,.
\end{align}
Therefore, the dynamics of the system is encapsulated in Eq.~\eqref{eq:SchrodingerOneD}, equipped with the potential given in Eq.~\eqref{eq:PotentialSpinNewton}. 
We compare this result against Eq.~\eqref{Helmholtz} with the gravitational potential of a point mass lens, namely $U(r) = - M/r$. 
In order, in Eq.~\eqref{eq:PotentialSpinNewton}  we recognize the gravitational potential of the same lens (first term) and the angular momentum contributions. The latter is explicitly present in Eq.~\eqref{eq:PotentialSpinNewton} because we have performed a separation of variables and decomposed the field on a spherical harmonics basis. A similar term can be made manifest also in Eq.~\eqref{Helmholtz} by performing the same manipulations.
Furthermore the potential above, contains two spin-dependent terms, absent in the previous description valid for scalar waves. 
Among these two, we recognize in the first a contribution to the angular momentum, building up to the total angular momentum $J^2 = L^2 + S^2$.  
Even though the general theory has these additional spin-dependent contributions, we claim that these two are negligible in the high frequency limit. This is already evident from Eq.~\eqref{eq:PotentialSpinNewton} since the spin dependent terms are always suppressed by different powers of the frequency. Indeed, $s$ is bounded to be smaller than $2$, while the angular momentum can take arbitrarily large numbers to keep $l(l+1)/ (\omega^2 r^2)$ potentially finite even in the $\omega \gg 1$ limit.
Note that this behavior might change close to the horizon. For a non-rotating BH, the potential~\eqref{eq:PotentialBH} can be written as
\begin{align}\label{eq:VslmBH_Schwarzschild}
    V^{s}_{\ell m}(\omega, r) = \frac{4 M r^2 (M - r)}{(r^2 - 2 M r)^2} + \frac{1}{\omega} \frac{2 i s (r - 3 M)}{(r - 2 M)^2} + \frac{1}{\omega^2} \left( \frac{l(l+1) (r^2 - 2 M r) + M (s^2 - 1)}{r^2 (r - 2 M)^2}\right) \,,
\end{align}
where we organized the terms in powers of $1/\omega$.
Sufficiently far away from the horizon, where Newtonian theory is valid, the spin-dependent contributions vanish in the limit $\omega \to \infty$. However, the horizon corrections can boost the spin dependent terms (second and fourth).
This fact also tells us that neglecting spin effects in the diffraction integral, as described in Section~\ref{sec:nakamura_approximations}, is not an inconsistent assumption, since this relies on the eikonal and paraxial approximations (which pose a lower bound on $\omega$) and it is derived in the weak field limit. 
We point out that this fact has a clear interpretation in the context outlined in the previous Sections: in the limit where a dispersion relation can be assigned to a wave, these can be described in terms of effective particles propagating on geodesics of the background spacetime regardless of their spin.

\section{Conclusions}

Gravitational lensing of gravitational waves is a rich field of research at the intersection of astrophysics and cosmology. While geometric optics offers a powerful tool for analyzing gravitational waves in the high frequency regime ($\lambda \ll {\cal R}_s$), recent years have seen a growing interest in the opposite optical regime, the wave optics one.
Its current description relies on the diffraction integral, which we summarized in Section~\ref{sec:nakamura_approximations}. While this derivation has the advantage to make manifest the interpretation of the field in terms of the associated particle, and smoothly transition to geometric optics in the high-frequency limit, it is  largely based on the {\it paraxial} and {\it eikonal} approximations, effectively posing a lower frequency limit on the validity range of such a description. Moving beyond these approximations is crucial. It not only deepens our theoretical understanding of the behavior of gravitational waves in curved spacetimes, but it also becomes vital as efforts intensify to probe the Universe's matter content using low-frequency gravitational lensing.

In this work, we proposed a generalization of the diffraction integral based on the introduction of a fifth coordinate, called {\it proper time} (see Section~\ref{sec:ProperTimeScalar}). 
The introduction of this new variable allows us to seamlessly transition from the standard Helmholtz equation for wave propagation, to a Schr\"odinger-like one, without the need of any approximation. 
This procedure led us directly to the {\it Feynman-Fradkin} exact representation of the Green function in Eq.~\eqref{G:path_integral} for a scalar field propagating through a spacetime containing a gravitational lens in the Newtonian regime. 
Just as in the diffraction integral, also in our formalism the wave's frequency plays the role of $1 / \hbar$, allowing us to set up the Hamilton analogy and compare wave effects to the quantum ones. 
The worldline representation of the Green function, $\tilde G_\omega ({\bf x}_f, {\bf x}_i, \tau)$, is once again constructed out of particle-like path integrals over the paths, rather than one over different field's configurations, making possible a direct link to the laws of high-frequency optics, since the proper time is removed upon integration at the end.

In Section~\ref{sec:HFresults}, we investigated the high-frequency limit ($\omega \to \infty$) using three approaches. First, we analyze the WKB form of the propagator through saddle point approximations for both proper time and trajectories. This recovers the geometric optics (GO) ansatz in Eq.~\eqref{eq:GWKB}.
Next, aiming to obtain the diffraction integral and eliminate the proper time integration while preserving the path integral over trajectories, we assume the ability to interchange the integration order within the Green's function. This is an inexact procedure, but established as a good approximation in the high-frequency regime~\cite{Garrod_1966,Gutzwiller_1967}.
Following the order swap, the proper time integral is evaluated in two ways. One way involves strictly fixing for the proper time $\tau$ at its classical configuration satisfying the equations of motion~\eqref{Constraint_refractive}. The other way implements a saddle point approximation that considers second-order deviations from the classical value $\sim (\tau - \tau_{cl})^2$. The first approach yields the diffraction integral, demonstrating how our formalism encompasses the standard approach. The second approach leads to the well-known {\it Feynman-Garrod} direct representation of the propagator.

After assessing the high-frequency behavior, we implemented a perturbative approach in Section~\ref{sec:perturbative_approach} with respect to the associated potential, considering the limit $\omega \alpha \ll 1$. We demonstrated how the expansion series can be organized in terms of waves propagating freely, except for an always increasing number of scattering events. This is equivalent to organizing the perturbation series such that terms are always of higher orders in the Born approximation. In the ideal limit of $n \to \infty$,  where $n$ represents the perturbation order, the wave scatters with every point of the gravitational potential. Consequently, there is no more free propagation within the region where $V \neq 0$. This observation suggests the possibility of an exact resummation of the perturbation scheme. Indeed, it leads us to the Dyson equation for the wave's propagator, where the gravitational potential clearly plays the role of proper self-energy, as also suggested by the Hamiltonian derivation of Section~\ref{sec:Hamilton}.
Our path integral description also allows for the introduction of a generating functional, $Z$, implemented through the introduction of a current, $Z$, coupled to the path ${\bf x}(\tau')$.

Subsequently, we explored two explicit applications of our formalism. Firstly,in Section~\ref{sec:Yukawa}, we present explicit computations for the first-order Green function in the case of a Coulomb potential, providing an estimate of the modifications to the free Green function, or the distance to the source, induced by the presence of the lens in the $\alpha \omega \ll 1$ limit.
Our results demonstrate the characteristic divergence in the forward limit of a long-range interaction, alongside the expected attenuation of the effect as we focus in the diffractive limit $\omega M G \to 0$ or $\lambda \gg {\cal R}_s$, where $\lambda$ represents the wavelength of the wave and ${\cal R}_s$ denotes the Schwarzschild radius of the lens. Then, in Section~\ref{sec:MassiveScalarField}, we apply the formalism to the case of a massive scalar field demonstrating that our method can be straightforwardly implemented by performing a frequency redefinition.

In Section~\ref{sec:spin2 for spheric lenses}, we addressed another limitation of the diffraction integral: its neglect of polarization effects, as it is valid only for scalar waves. Previous treatments of the diffractive regime often assume the conservation of polarization associated with gravitational waves during propagation, potentially overlooking valuable insights.

To explore polarization effects, we specialize the lens model to BHs with axisymmetry, and consider a Kerr background. This choice streamlines the derivation, allowing us to leverage results from the established field of BH perturbation theory. For perturbations of various spin ($s = 0, 1/2, 1, 2$), the equations of motion on the Kerr background are well-known in the literature and are described by the Teukolsky equation, in terms of Newman-Penrose scalar.
After separating the radial and angular components, we reformulate the radial part of the Teukolsky equation as a one-dimensional Helmholtz equation, with a potential dependent on frequency, spin, and angular momentum. This formulation enables us to derive the Feynman-Fradkin representation of the propagator for higher spin fields.
The role of spin has then been investigated by specializing to the case of non-rotating BH by setting $a = 0$, obtaining two main equations. One which identifies spin-dependent effects within the same regime of validity as the initial Helmholtz equation evaluated for a spherically symmetric lens. Particularly, this result shows that one of the two spin-dependent contributions, $s(s+1)/(\omega^2 r^2)$, contributes to the total angular momentum, with both contributions being suppressed by higher powers of frequency.  The other equation demonstrates that spin contributions can become significant in the near-horizon region, where the largeness of $\omega$ can be offset the vanishing denominators proportional to $r - 2 M$. However, the diffraction integral is valid only for Newtonian lenses, thus it cannot account for near-horizon contributions. These results confirm that neglecting polarization effects in the derivation of the diffraction integral aligns with the underlying assumptions of the eikonal limit (with the frequency lower bound from Eq.~\eqref{eq:WKB_Assumption}) and the weak field $\alpha \ll 1$ approximation.

Applying our formalism to specific lensing scenario and assessing its potentiality into investigating the properties of the lenses, will be object of future works.

\acknowledgments
We would like to thank J. Feldbrugge, D. Bertacca, M.  Zumalacárregui, G. Gradenigo for their useful discussions and comments. G.B. thanks A. Sesana for his support.
S.M and N.B acknowledge the funding from the MUR Departments of Excellence grant "Quantum Frontiers".
A.G. is supported by funds provided by the Center for Particle Cosmology at the University of Pennsylvania. 
A.R. acknowledges financial support from the Supporting TAlent in ReSearch@University of Padova (STARS@UNIPD) for the project “Constraining Cosmology and Astrophysics with Gravitational Waves, Cosmic Microwave Background and Large-Scale Structure cross-correlations”. 

\appendix

\section{Free propagator computations}\label{app:FreeProp}

In this appendix we provide the explicit steps to compute the free propagator $G_\omega ({\bf x}_f,{\bf x}_i)$ following~\cite{corradini2021spinning}. 
We start from its worldline representation, namely Eq.~\eqref{eq:PerturbationTheoryFirst}, where 
\begin{equation}
    S_0 \equiv \int^\tau_0 d \tau' \, \frac14 \left( \frac{d{\bf x}(\tau')}{d \tau'}\right)^2\,.
\end{equation}
The first step to perform the path integral, is to redefine the path and consider a perturbation of it from the classical solution, i.e. we define ${\bf x}(\tau') = {\bf x}_{cl}(\tau') + {\bf q}(\tau')$. Since the total and the classical paths satisfy the same boundary conditions, we have that ${\bf q}(\tau' = \tau) = {\bf q}(\tau' = 0)  = 0$.
The classical trajectory ${\bf x}_{cl}(\tau)$ satisfies the Euler-Lagrange equation of motion, descending from the free action $S_0$, namely $ \ddot{\bf x}_{cl} = 0$. This means 
\begin{equation}
   \dot{\bf x}_{cl} = \text{const}\,, \qquad \to \qquad \dot{\bf x}_{cl} = \frac{{\bf x}_f - {\bf x}_i}{\tau}
\end{equation}
as the motion between ${\bf x}_f$ and ${\bf x}_f$ occurs in a time period of $\tau$ at constant speed.
This allows us to write the free action as 
\begin{equation}
    S_0  = \frac{|{\bf x}_f - {\bf x}_i|^2}{ 4 \tau^2} + \int^\tau_0 d \tau' \frac{\dot {\bf q}^2}{4}\,.
\end{equation}
We evaluate the path integral of Eq.~\eqref{eq:PerturbationTheoryFirst} as 
\begin{align}
     G^{(0)}_\omega ({\bf x}_f,{\bf x}_i,\tau) &= \int^{{\bf x}(\tau' = \tau) = {\bf x}_f}_{{\bf x}(\tau' = 0) = {\bf x}_i} \mathcal{D}{\bf x}(\tau^\prime)\, \exp\left[i\omega \int_0^\tau d\tau^\prime \frac{\dot {\bf x}^2}{4}\right] = \nonumber \\
     &= e^{\frac{i \omega}{4 \tau}|{\bf x}_f - {\bf x}_i|^2 }\int^{{\bf q}(\tau' = \tau) = 0}_{{\bf q}(\tau' = 0) = 0} \mathcal{D}{\bf q}(\tau^\prime)\, \exp\left[i\omega \int_0^\tau d\tau^\prime \frac{\dot {\bf q}^2}{4}\right] = \nonumber \\
     &= \left( \frac{\omega}{4 \pi i \tau} \right)^{3/2} e^{\frac{i \omega}{4 \tau}|{\bf x}_f - {\bf x}_i|^2 }\,.
\end{align}
Finally, the propagator is obtained by performing the proper time integral
\begin{align}
    G^{(0)}_\omega ({\bf x}_f,{\bf x}_i) &= -\frac{i}{\omega} \int^{+\infty}_0 \, d \tau \, \left( \frac{\omega}{4 \pi i \tau} \right)^{3/2} e^{\frac{i \omega}{4 \tau}|{\bf x}_f - {\bf x}_i|^2 + i \omega \tau} = - \frac{1}{4 \pi} \frac{e^{i \omega |{\bf x}_f - {\bf x}_i|}}{|{\bf x}_f - {\bf x}_i|}\,,
\end{align}
so that, in real time, we have 
\begin{equation}
     G^{(0)}({\bf x}_f,{\bf x}_i, t) = - \frac{1}{4 \pi} \int \frac{d \omega}{2 \pi } \frac{e^{-i \omega (t- |{\bf x}_f - {\bf x}_i|)}}{|{\bf x}_f - {\bf x}_i|} = - \frac{1}{8 \pi^2} \frac{\delta(t - |{\bf x}_f - {\bf x}_i|)}{|{\bf x}_f - {\bf x}_i|}\,,
\end{equation}
displaying the correct behavior (up to a normalization factor). 
To obtain the momentum space Green function, we perform the 3D Fourier transform 
\begin{align}
    G^{(0)}_\omega({\bf p}) &= - \frac{1}{4 \pi} \int d^3 x \, \frac{e^{i \omega |{\bf x}_f - {\bf x}_i|- i {\bf p} \cdot ({\bf x}_f - {\bf x}_i)}}{|{\bf x}_f - {\bf x}_i|} = \nonumber \\
    &= -\frac{1}{4 \pi} \int^{2 \pi}_0 d \varphi \int^{1}_{-1} d \mu \int^{+\infty}_0 d x \: x^2 \frac{e^{i \omega x - i p x \mu}}{x} = \nonumber \\
    &= \frac{1}{p^2 - \omega^2}\,.
\end{align}

\section{Coulomb potential computations}\label{app:Coulomb}
In this Appendix we provide the explicit computations of Section~\ref{sec:Yukawa}.
In particular, we illustrate the derivation of Eqs.~\eqref{eq:FirstGreenProperTimeYukawa} and ~\eqref{eq:FirstGreenYukawa}. 
First we point out that the integral in Eq.~\eqref{eq:IntegralGsquare} can be easily performed using (see~\cite{Feynman:100771})
\begin{equation}
  \int^T_0  d \tau \: \frac{e^{- \frac{a}{T - \tau} - \frac{b}{\tau}}}{ \left[ \tau (T - \tau)\right]^{3/2}}  = \sqrt{\frac{\pi}{T^3}} \frac{\sqrt{a} + \sqrt{b}}{\sqrt{a b}} \: e^{\frac{-\left( \sqrt{a} + \sqrt{b}\right)^2}{T}}\,.
\end{equation}
Then we show the derivation of Eq.~\eqref{eq:FirstGreenProperTimeYukawa}, namely the worldline form of the first order propagator. 
As explained in the main text, we consider a scattering situation in which the wave's source and observer are located very far away form the lens. 
In practice, this means considering the approximation in Eq.~\eqref{eq:SOfarfromL} for $|{\bf x}_O - {\bf x}^\star|$ and $|{\bf x}^\star - {\bf x}_S|$ and approximate the result of Eq.~\eqref{eq:IntegralGsquare} considering the lowest order in $r^\star/ r_S$ and $r^\star/ r_O$ in the multiplicative parenthesis, while retaining the first order correction in the phase of the exponential. 
Therefore, we consider the approximation of Eq.~\eqref{eq:IntegralGsquare}
\begin{align}
 \int^\tau_0 \, d \tau' \, &\tilde G^{(0)}_\omega({\bf x}_O,{\bf x}^\star, \tau-\tau' ) V({\bf x}^\star) \tilde G^{(0)}_\omega({\bf x}^\star,{\bf x}_S, \tau') = \nonumber \\
 &\approx \left( \frac{\omega}{4 i \pi}\right)^{5/2} \tau^{-3/2} \left( \frac{1}{r_O} + \frac{1}{r_S}\right) e^{\frac{i \omega}{4 \tau} (r_S + r_O)^2} e^{\frac{i \omega}{2 \tau} {\bf x}^\star \cdot {\bf R}} \: V({\bf x}^\star)
\end{align}
where we have introduced the vector ${\bf R} \equiv (r_S + r_O) (\boldsymbol{\theta}_S - \boldsymbol{\theta}_O)$.
Now we compute the integral over the interaction point ${\bf x}^\star$ as prescribed in Eq.~\eqref{eq:proper_time_second_order}.
The steps to obtain Eq.~\eqref{eq:FirstGreenProperTimeYukawa} are
\begin{align}
    \tilde G^{(1)}_\omega({\bf x}_O,{\bf x}_S, \tau)&= \left( \frac{\omega}{4 i \pi}\right)^{5/2} \tau^{-3/2} \left( \frac{1}{r_O} + \frac{1}{r_S}\right) e^{\frac{i \omega}{4 \tau} (r_S + r_O)^2} \int d^3 {\bf x}^\star e^{\frac{i \omega}{2 \tau} {\bf x}^\star \cdot {\bf R}} \: V({\bf x}^\star) = \nonumber \\
   &= -4 \alpha MG \left( \frac{\omega}{4 i \pi}\right)^{5/2} \tau^{-3/2} \left( \frac{1}{r_O} + \frac{1}{r_S}\right) e^{\frac{i \omega}{4 \tau} (r_S + r_O)^2} \times \nonumber \\
   &\qquad\times \int^{2 \pi}_0 d \varphi^\star \int^1_{-1} d \mu^\star \int^{+\infty}_0 d r^\star \: {r^\star}^2 \: \frac{e^{\frac{i \omega}{2 \tau} R r^\star \mu^\star - r^\star/a}}{r^\star} = \nonumber \\
   &=- 8 \pi \alpha MG \left( \frac{\omega}{4 i \pi}\right)^{5/2} \tau^{-3/2} \left( \frac{1}{r_O} + \frac{1}{r_S}\right) e^{\frac{i \omega}{4 \tau} (r_S + r_O)^2} \left( \frac{2 \tau}{i \omega R}  \right) \times \nonumber \\
   &\qquad\times \int^{+\infty}_0 d r^\star \: \left[ e^{r^\star \left[\frac{i \omega}{2 \tau} R - \frac{1}{a}\right]} - e^{-r^\star \left[\frac{i \omega}{2 \tau} R + \frac{1}{a}\right]}\right]= \nonumber \\
   &= - 8 \pi \alpha MG \left( \frac{\omega}{4 i \pi}\right)^{5/2} \tau^{-3/2} \left( \frac{1}{r_O} + \frac{1}{r_S}\right) e^{\frac{i \omega}{4 \tau} (r_S + r_O)^2} \left( \frac{2 \tau}{i \omega R}  \right) \times \nonumber \\
   &\qquad\times \left[ \frac{e^{r^\star \left[\frac{i \omega}{2 \tau} R - \frac{1}{a}\right]} }{\frac{i \omega}{2 \tau} R - \frac{1}{a}} + \frac{e^{-r^\star \left[\frac{i \omega}{2 \tau} R + \frac{1}{a}\right]}}{\frac{i \omega}{2 \tau} R + \frac{1}{a}}\right]^{+\infty}_0= \nonumber \\
   &= 16 \pi \alpha MG \left( \frac{\omega}{4 i \pi}\right)^{5/2}  \left( \frac{1}{r_O} + \frac{1}{r_S}\right)  \times   \frac{\tau^{-3/2}\, e^{\frac{i \omega}{4 \tau} (r_S + r_O)^2} }{\left( \frac{i \omega R}{2 \tau}\right)^2 - \frac{1}{a^2}} \,,
\end{align}
where $\mu^\star$ is the cosine of the angle between ${\bf R}$ and ${\bf x}^\star$.
We use this result to perform the integral over the proper time and obtain the first order propagator. 
That is, we explicitly show the integration that leads to Eq.~\eqref{eq:FirstGreenYukawa}, where we have considered a long range interaction and send $a \to \infty$. These are
\begin{align}
    G^{(1)}_\omega({\bf x}_O,{\bf x}_S) &\equiv -\frac{i}{\omega} \, \, \int^{+\infty}_0 d \tau \: e^{i \omega \tau} \: \tilde G^{(1)}_\omega({\bf x},{\bf x}_0, \tau) = \nonumber \\
     &= 16 \pi \alpha MG  \left(- \frac{i}{\omega} \right)   \left( \frac{\omega}{4 i \pi}\right)^{5/2} \left( \frac{1}{r_O} + \frac{1}{r_S}\right) \left( \frac{-4}{ \omega^2 R^2} \right)  \int^{+\infty}_0 d \tau \:  \sqrt{\tau }\, e^{\frac{i \omega}{4 \tau} (r_S + r_O)^2 + i \omega \tau}= \nonumber \\
     &= 16 \pi \alpha MG  \left(- \frac{i}{\omega} \right)   \left( \frac{\omega}{4 i \pi}\right)^{5/2} \left( \frac{1}{r_O} + \frac{1}{r_S}\right) \left( \frac{-4}{ \omega^2 R^2} \right) \times\nonumber \\
     &\qquad\times  \left( \frac{- \sqrt{- i \pi \omega}}{2 \omega^2}\right) e^{i \omega(r_S + r_O) } (1 - i \omega(r_S + r_O))= \nonumber \\
     &= \frac{ \alpha MG}{\pi \omega^2} \left( \frac{1}{r_O} + \frac{1}{r_S}\right)  \frac{e^{i \omega (r_S + r_O)} (1 - i \omega (r_S + r_O))}{R^2}\,.
\end{align}
Note that, in order to perform the integral, we regularized it by analytically continuing $\omega$ in the complex plane, with the prescription $\omega \to \omega + i \epsilon$, and the then took the limit $\epsilon \to 0$.

\bibliography{bibliography}

\providecommand{\href}[2]{#2}\begingroup\raggedright\begin{thebibliography}{100}

\bibitem{Takahashi:2003ix}
R.~Takahashi and T.~Nakamura, {\it {Wave effects in gravitational lensing of gravitational waves from chirping binaries}},  {\em Astrophys. J.} {\bf 595} (2003) 1039--1051, [\href{http://arxiv.org/abs/astro-ph/0305055}{{\tt astro-ph/0305055}}].

\bibitem{Gao:2021sxw}
Z.~Gao, X.~Chen, Y.-M. Hu, J.-D. Zhang, and S.-J. Huang, {\it {A higher probability of detecting lensed supermassive black hole binaries by LISA}},  {\em Mon. Not. Roy. Astron. Soc.} {\bf 512} (2022), no.~1 1--10, [\href{http://arxiv.org/abs/2102.10295}{{\tt arXiv:2102.10295}}].

\bibitem{Dai:2018enj}
L.~Dai, S.-S. Li, B.~Zackay, S.~Mao, and Y.~Lu, {\it {Detecting Lensing-Induced Diffraction in Astrophysical Gravitational Waves}},  {\em Phys. Rev. D} {\bf 98} (2018), no.~10 104029, [\href{http://arxiv.org/abs/1810.00003}{{\tt arXiv:1810.00003}}].

\bibitem{Cheung:2020okf}
M.~H.~Y. Cheung, J.~Gais, O.~A. Hannuksela, and T.~G.~F. Li, {\it {Stellar-mass microlensing of gravitational waves}},  {\em Mon. Not. Roy. Astron. Soc.} {\bf 503} (2021), no.~3 3326--3336, [\href{http://arxiv.org/abs/2012.07800}{{\tt arXiv:2012.07800}}].

\bibitem{Reitze:2019iox}
D.~Reitze et~al., {\it {Cosmic Explorer: The U.S. Contribution to Gravitational-Wave Astronomy beyond LIGO}},  {\em Bull. Am. Astron. Soc.} {\bf 51} (2019), no.~7 035, [\href{http://arxiv.org/abs/1907.04833}{{\tt arXiv:1907.04833}}].

\bibitem{Punturo:2010zz}
M.~Punturo et~al., {\it {The Einstein Telescope: A third-generation gravitational wave observatory}},  {\em Class. Quant. Grav.} {\bf 27} (2010) 194002.

\bibitem{Branchesi:2023mws}
M.~Branchesi et~al., {\it {Science with the Einstein Telescope: a comparison of different designs}},  {\em JCAP} {\bf 07} (2023) 068, [\href{http://arxiv.org/abs/2303.15923}{{\tt arXiv:2303.15923}}].

\bibitem{LISAConsortiumWaveformWorkingGroup:2023arg}
{\bf LISA Consortium Waveform Working Group} Collaboration, N.~Afshordi et~al., {\it {Waveform Modelling for the Laser Interferometer Space Antenna}},  \href{http://arxiv.org/abs/2311.01300}{{\tt arXiv:2311.01300}}.

\bibitem{Caliskan:2022hbu}
M.~\c{C}al\i{}\c{s}kan, L.~Ji, R.~Cotesta, E.~Berti, M.~Kamionkowski, and S.~Marsat, {\it {Observability of lensing of gravitational waves from massive black hole binaries with LISA}},  {\em Phys. Rev. D} {\bf 107} (2023), no.~4 043029, [\href{http://arxiv.org/abs/2206.02803}{{\tt arXiv:2206.02803}}].

\bibitem{Tambalo:2022wlm}
G.~Tambalo, M.~Zumalac\'arregui, L.~Dai, and M.~H.-Y. Cheung, {\it {Gravitational wave lensing as a probe of halo properties and dark matter}},  {\em Phys. Rev. D} {\bf 108} (2023), no.~10 103529, [\href{http://arxiv.org/abs/2212.11960}{{\tt arXiv:2212.11960}}].

\bibitem{Lai_2018}
K.-H. Lai, O.~A. Hannuksela, A.~Herrera-Martín, J.~M. Diego, T.~Broadhurst, and T.~G. Li, {\it Discovering intermediate-mass black hole lenses through gravitational wave lensing},  {\em Physical Review D} {\bf 98} (Oct., 2018).

\bibitem{Savastano:2023spl}
S.~Savastano, G.~Tambalo, H.~Villarrubia-Rojo, and M.~Zumalacarregui, {\it {Weakly lensed gravitational waves: Probing cosmic structures with wave-optics features}},  {\em Phys. Rev. D} {\bf 108} (2023), no.~10 103532, [\href{http://arxiv.org/abs/2306.05282}{{\tt arXiv:2306.05282}}].

\bibitem{Yeung:2021chy}
S.~M.~C. Yeung, M.~H.~Y. Cheung, E.~Seo, J.~A.~J. Gais, O.~A. Hannuksela, and T.~G.~F. Li, {\it {Detectability of microlensed gravitational waves}},  {\em Mon. Not. Roy. Astron. Soc.} {\bf 526} (2023), no.~2 2230--2240, [\href{http://arxiv.org/abs/2112.07635}{{\tt arXiv:2112.07635}}].

\bibitem{Diego:2019lcd}
J.~M. Diego, O.~A. Hannuksela, P.~L. Kelly, T.~Broadhurst, K.~Kim, T.~G.~F. Li, G.~F. Smoot, and G.~Pagano, {\it {Observational signatures of microlensing in gravitational waves at LIGO/Virgo frequencies}},  {\em Astron. Astrophys.} {\bf 627} (2019) A130, [\href{http://arxiv.org/abs/1903.04513}{{\tt arXiv:1903.04513}}].

\bibitem{Guo:2022dre}
X.~Guo and Y.~Lu, {\it {Probing the nature of dark matter via gravitational waves lensed by small dark matter halos}},  {\em Phys. Rev. D} {\bf 106} (2022), no.~2 023018, [\href{http://arxiv.org/abs/2207.00325}{{\tt arXiv:2207.00325}}].

\bibitem{Jung:2017flg}
S.~Jung and C.~S. Shin, {\it {Gravitational-Wave Fringes at LIGO: Detecting Compact Dark Matter by Gravitational Lensing}},  {\em Phys. Rev. Lett.} {\bf 122} (2019), no.~4 041103, [\href{http://arxiv.org/abs/1712.01396}{{\tt arXiv:1712.01396}}].

\bibitem{Fairbairn:2022xln}
M.~Fairbairn, J.~Urrutia, and V.~Vaskonen, {\it {Microlensing of gravitational waves by dark matter structures}},  {\em JCAP} {\bf 07} (2023) 007, [\href{http://arxiv.org/abs/2210.13436}{{\tt arXiv:2210.13436}}].

\bibitem{Urrutia:2024pos}
J.~Urrutia and V.~Vaskonen, {\it {The dark timbre of gravitational waves}},  \href{http://arxiv.org/abs/2402.16849}{{\tt arXiv:2402.16849}}.

\bibitem{Urrutia:2023mtk}
J.~Urrutia, V.~Vaskonen, and H.~Veerm\"ae, {\it {Gravitational wave microlensing by dressed primordial black holes}},  {\em Phys. Rev. D} {\bf 108} (2023), no.~2 023507, [\href{http://arxiv.org/abs/2303.17601}{{\tt arXiv:2303.17601}}].

\bibitem{Urrutia:2021qak}
J.~Urrutia and V.~Vaskonen, {\it {Lensing of gravitational waves as a probe of compact dark matter}},  {\em Mon. Not. Roy. Astron. Soc.} {\bf 509} (2021), no.~1 1358--1365, [\href{http://arxiv.org/abs/2109.03213}{{\tt arXiv:2109.03213}}].

\bibitem{Sugiyama:2019dgt}
S.~Sugiyama, T.~Kurita, and M.~Takada, {\it {On the wave optics effect on primordial black hole constraints from optical microlensing search}},  {\em Mon. Not. Roy. Astron. Soc.} {\bf 493} (2020), no.~3 3632--3641, [\href{http://arxiv.org/abs/1905.06066}{{\tt arXiv:1905.06066}}].

\bibitem{Diego:2019rzc}
J.~M. Diego, {\it {Constraining the abundance of primordial black holes with gravitational lensing of gravitational waves at LIGO frequencies}},  {\em Phys. Rev. D} {\bf 101} (2020), no.~12 123512, [\href{http://arxiv.org/abs/1911.05736}{{\tt arXiv:1911.05736}}].

\bibitem{Oguri:2020ldf}
M.~Oguri and R.~Takahashi, {\it {Probing Dark Low-mass Halos and Primordial Black Holes with Frequency-dependent Gravitational Lensing Dispersions of Gravitational Waves}},  {\em Astrophys. J.} {\bf 901} (2020), no.~1 58, [\href{http://arxiv.org/abs/2007.01936}{{\tt arXiv:2007.01936}}].

\bibitem{GilChoi:2023ahp}
H.~Gil~Choi, S.~Jung, P.~Lu, and V.~Takhistov, {\it {Co-Existence Test of Primordial Black Holes and Particle Dark Matter}},  \href{http://arxiv.org/abs/2311.17829}{{\tt arXiv:2311.17829}}.

\bibitem{Basak:2021ten}
S.~Basak, A.~Ganguly, K.~Haris, S.~Kapadia, A.~K. Mehta, and P.~Ajith, {\it {Constraints on Compact Dark Matter from Gravitational Wave Microlensing}},  {\em Astrophys. J.} {\bf 926} (2022), no.~2 L28, [\href{http://arxiv.org/abs/2109.06456}{{\tt arXiv:2109.06456}}].

\bibitem{Nakamura1999WaveOI}
T.~T. Nakamura and S.~Deguchi, {\it Wave optics in gravitational lensing},  {\em Progress of Theoretical Physics Supplement} {\bf 133} (1999) 137--153.

\bibitem{Nakamura:1997sw}
T.~T. Nakamura, {\it {Gravitational lensing of gravitational waves from inspiraling binaries by a point mass lens}},  {\em Phys. Rev. Lett.} {\bf 80} (1998) 1138--1141.

\bibitem{Takahashi:2004mc}
R.~Takahashi, {\it {Quasigeometrical optics approximation in gravitational lensing}},  {\em Astron. Astrophys.} {\bf 423} (2004) 787--792, [\href{http://arxiv.org/abs/astro-ph/0402165}{{\tt astro-ph/0402165}}].

\bibitem{Takahashi:2005sxa}
R.~Takahashi, T.~Suyama, and S.~Michikoshi, {\it {Scattering of gravitational waves by the weak gravitational fields of lens objects}},  {\em Astron. Astrophys.} {\bf 438} (2005) L5, [\href{http://arxiv.org/abs/astro-ph/0503343}{{\tt astro-ph/0503343}}].

\bibitem{Takahashi:2005ug}
R.~Takahashi, {\it {Amplitude and phase fluctuations for gravitational waves propagating through inhomogeneous mass distribution in the universe}},  {\em Astrophys. J.} {\bf 644} (2006) 80--85, [\href{http://arxiv.org/abs/astro-ph/0511517}{{\tt astro-ph/0511517}}].

\bibitem{Schneider1992}
P.~Schneider, J.~Ehlers, and E.~E. Falco, {\em Gravitational Lenses}.
\newblock Springer Berlin Heidelberg, 1992.

\bibitem{Born_Wolf_Bhatia_Clemmow_Gabor_Stokes_Taylor_Wayman_Wilcock_1999}
M.~Born, E.~Wolf, A.~B. Bhatia, P.~C. Clemmow, D.~Gabor, A.~R. Stokes, A.~M. Taylor, P.~A. Wayman, and W.~L. Wilcock, {\em Principles of Optics: Electromagnetic Theory of Propagation, Interference and Diffraction of Light}.
\newblock Cambridge University Press, 7~ed., 1999.

\bibitem{leung2023wave}
C.~Leung, D.~Jow, P.~Saha, L.~Dai, M.~Oguri, and L.~V.~E. Koopmans, {\it Wave mechanics, interference, and decoherence in strong gravitational lensing},  2023.

\bibitem{goldstein2002classical}
H.~Goldstein, C.~Poole, and J.~Safko, {\em Classical Mechanics}.
\newblock Addison Wesley, 2002.

\bibitem{Masoliver_2009}
J.~Masoliver and A.~Ros, {\it From classical to quantum mechanics through optics},  {\em European Journal of Physics} {\bf 31} (Nov., 2009) 171–192.

\bibitem{Feldbrugge:2019fjs}
J.~Feldbrugge, U.-L. Pen, and N.~Turok, {\it {Oscillatory path integrals for radio astronomy}},  {\em Annals Phys.} {\bf 451} (2023) 169255, [\href{http://arxiv.org/abs/1909.04632}{{\tt arXiv:1909.04632}}].

\bibitem{Fishman1984}
L.~Fishman and J.~J. McCoy, {\it {Derivation and application of extended parabolic wave theories. II. Path integral representations}},  {\em Journal of Mathematical Physics} {\bf 25} (02, 1984) 297--308, [\href{http://arxiv.org/abs/https://pubs.aip.org/aip/jmp/article-pdf/25/2/297/8153852/297\_1\_online.pdf}{{\tt https://pubs.aip.org/aip/jmp/article-pdf/25/2/297/8153852/297\_1\_online.pdf}}].

\bibitem{Fishman_2006}
L.~Fishman, {\it Helmholtz path integrals},  {\em AIP Conference Proceedings} {\bf 834} (05, 2006).

\bibitem{Schulman1981TechniquesAA}
L.~S. Schulman and C.~DeWitt-Morette, {\it Techniques and applications of path integration},  1981.

\bibitem{Eve:PI-wavetheory}
M.~Eve, {\it The use of path integrals in guided wave theory}, .

\bibitem{Alvarez_1998}
E.~T.~G. Alvarez and F.~H. Gaioli {\em Foundations of Physics} {\bf 28} (1998), no.~10 1529–1538.

\bibitem{space-timeQED}
R.~P. Feynman, {\it Space-time approach to quantum electrodynamics},  {\em Phys. Rev.} {\bf 76} (Sep, 1949) 769--789.

\bibitem{Nambu:1950rs}
Y.~Nambu, {\it {The use of the Proper Time in Quantum Electrodynamics}},  {\em Prog. Theor. Phys.} {\bf 5} (1950) 82--94.

\bibitem{RevModPhys.58.449}
S.~S. Schweber, {\it Feynman and the visualization of space-time processes},  {\em Rev. Mod. Phys.} {\bf 58} (Apr, 1986) 449--508.

\bibitem{schwinger}
J.~Schwinger, {\it On gauge invariance and vacuum polarization},  {\em Phys. Rev.} {\bf 82} (Jun, 1951) 664--679.

\bibitem{Fock:1937dy}
V.~Fock, {\it {Proper time in classical and quantum mechanics}},  {\em Phys. Z. Sowjetunion} {\bf 12} (1937) 404--425.

\bibitem{Strassler_1992}
M.~J. Strassler, {\it Field theory without feynman diagrams: One-loop effective actions},  {\em Nuclear Physics B} {\bf 385} (Oct., 1992) 145–184.

\bibitem{Bastianelli_2002}
F.~Bastianelli and A.~Zirotti, {\it Worldline formalism in a gravitational background},  {\em Nuclear Physics B} {\bf 642} (Oct., 2002) 372–388.

\bibitem{Bastianelli:2021nbs}
F.~Bastianelli, F.~Comberiati, and L.~de~la Cruz, {\it {Light bending from eikonal in worldline quantum field theory}},  {\em JHEP} {\bf 02} (2022) 209, [\href{http://arxiv.org/abs/2112.05013}{{\tt arXiv:2112.05013}}].

\bibitem{Bastianelli:2023oyz}
F.~Bastianelli and M.~D. Paciarini, {\it {Worldline path integrals for the graviton}},  \href{http://arxiv.org/abs/2305.06650}{{\tt arXiv:2305.06650}}.

\bibitem{corradini2021spinning}
O.~Corradini, C.~Schubert, J.~P. Edwards, and N.~Ahmadiniaz, {\it Spinning particles in quantum mechanics and quantum field theory},  2021.

\bibitem{Bonora:2018uwx}
L.~Bonora, M.~Cvitan, P.~Dominis~Prester, S.~Giaccari, M.~Pauli\v{s}i\'c, and T.~\v{S}temberga, {\it {Worldline quantization of field theory, effective actions and $L_\infty$ structure}},  {\em JHEP} {\bf 04} (2018) 095, [\href{http://arxiv.org/abs/1802.02968}{{\tt arXiv:1802.02968}}].

\bibitem{Edwards:2019eby}
J.~P. Edwards and C.~Schubert, {\it {Quantum mechanical path integrals in the first quantised approach to quantum field theory}},  12, 2019.
\newblock \href{http://arxiv.org/abs/1912.10004}{{\tt arXiv:1912.10004}}.

\bibitem{Edwards:2022dbd}
J.~P. Edwards, C.~M. Mata, and C.~Schubert, {\it {One-loop amplitudes in the worldline formalism}},  {\em Phys. Scripta} {\bf 97} (2022), no.~6 064002, [\href{http://arxiv.org/abs/2201.12457}{{\tt arXiv:2201.12457}}].

\bibitem{Feldbrugge:2017LQC}
J.~Feldbrugge, J.-L. Lehners, and N.~Turok, {\it {Lorentzian Quantum Cosmology}},  {\em Phys. Rev. D} {\bf 95} (2017), no.~10 103508, [\href{http://arxiv.org/abs/1703.02076}{{\tt arXiv:1703.02076}}].

\bibitem{Teitelboim:1981ua}
C.~Teitelboim, {\it {Quantum Mechanics of the Gravitational Field}},  {\em Phys. Rev. D} {\bf 25} (1982) 3159.

\bibitem{Teitelboim:1983fh}
C.~Teitelboim, {\it {Causality Versus Gauge Invariance in Quantum Gravity and Supergravity}},  {\em Phys. Rev. Lett.} {\bf 50} (1983) 705.

\bibitem{Teitelboim:1983fk}
C.~Teitelboim, {\it {The Proper Time Gauge in Quantum Theory of Gravitation}},  {\em Phys. Rev. D} {\bf 28} (1983) 297.

\bibitem{Gutzwiller2004}
M.~C. Gutzwiller, {\it {Phase‐Integral Approximation in Momentum Space and the Bound States of an Atom}},  {\em Journal of Mathematical Physics} {\bf 8} (12, 2004) 1979--2000, [\href{http://arxiv.org/abs/https://pubs.aip.org/aip/jmp/article-pdf/8/10/1979/8182197/1979\_1\_online.pdf}{{\tt https://pubs.aip.org/aip/jmp/article-pdf/8/10/1979/8182197/1979\_1\_online.pdf}}].

\bibitem{Gutzwiller_1967}
M.~C. Gutzwiller, {\it {Phase‐Integral Approximation in Momentum Space and the Bound States of an Atom}},  {\em Journal of Mathematical Physics} {\bf 8} (12, 2004) 1979--2000, [\href{http://arxiv.org/abs/https://pubs.aip.org/aip/jmp/article-pdf/8/10/1979/8182197/1979\_1\_online.pdf}{{\tt https://pubs.aip.org/aip/jmp/article-pdf/8/10/1979/8182197/1979\_1\_online.pdf}}].

\bibitem{Babington_crossing}
J.~Babington, {\it Ray–wave duality in classical optics: Crossing the feynman bridge},  {\em Optics Letters} {\bf 43} (11, 2018) 5591.

\bibitem{Babington_2021}
J.~Babington, {\it Ray-wave duality of electromagnetic fields: a feynman path integral approach to classical vectorial imaging},  {\em Journal of the Optical Society of America A} {\bf 38} (May, 2021) 817.

\bibitem{Hannay:pathlinking}
J.~H. Hannay, {\it Path-linking interpretation of kirchhoff diffraction},  {\em Proceedings: Mathematical and Physical Sciences} {\bf 450} (1995), no.~1938 51--65.

\bibitem{Schlottmann:1999}
R.~B. {Schlottmann}, {\it {A path integral formulation of acoustic wave propagation}},  {\em Geophysical Journal International} {\bf 137} (May, 1999) 353--363.

\bibitem{Palmer1979}
D.~R. Palmer, {\it {A path‐integral approach to the parabolic approximation. I}},  {\em The Journal of the Acoustical Society of America} {\bf 66} (09, 1979) 862--871, [\href{http://arxiv.org/abs/https://pubs.aip.org/asa/jasa/article-pdf/66/3/862/11940861/862\_1\_online.pdf}{{\tt https://pubs.aip.org/asa/jasa/article-pdf/66/3/862/11940861/862\_1\_online.pdf}}].

\bibitem{NANOGrav:2023gor}
{\bf NANOGrav} Collaboration, G.~Agazie et~al., {\it {The NANOGrav 15 yr Data Set: Evidence for a Gravitational-wave Background}},  {\em Astrophys. J. Lett.} {\bf 951} (2023), no.~1 L8, [\href{http://arxiv.org/abs/2306.16213}{{\tt arXiv:2306.16213}}].

\bibitem{EPTA:2023fyk}
{\bf EPTA, InPTA:} Collaboration, J.~Antoniadis et~al., {\it {The second data release from the European Pulsar Timing Array - III. Search for gravitational wave signals}},  {\em Astron. Astrophys.} {\bf 678} (2023) A50, [\href{http://arxiv.org/abs/2306.16214}{{\tt arXiv:2306.16214}}].

\bibitem{Reardon:2023gzh}
D.~J. Reardon et~al., {\it {Search for an Isotropic Gravitational-wave Background with the Parkes Pulsar Timing Array}},  {\em Astrophys. J. Lett.} {\bf 951} (2023), no.~1 L6, [\href{http://arxiv.org/abs/2306.16215}{{\tt arXiv:2306.16215}}].

\bibitem{Xu:2023wog}
H.~Xu et~al., {\it {Searching for the Nano-Hertz Stochastic Gravitational Wave Background with the Chinese Pulsar Timing Array Data Release I}},  {\em Res. Astron. Astrophys.} {\bf 23} (2023), no.~7 075024, [\href{http://arxiv.org/abs/2306.16216}{{\tt arXiv:2306.16216}}].

\bibitem{Figueroa:2023zhu}
D.~G. Figueroa, M.~Pieroni, A.~Ricciardone, and P.~Simakachorn, {\it {Cosmological Background Interpretation of Pulsar Timing Array Data}},  {\em Phys. Rev. Lett.} {\bf 132} (2024), no.~17 171002, [\href{http://arxiv.org/abs/2307.02399}{{\tt arXiv:2307.02399}}].

\bibitem{Contaldi:2016koz}
C.~R. Contaldi, {\it {Anisotropies of Gravitational Wave Backgrounds: A Line Of Sight Approach}},  {\em Phys. Lett. B} {\bf 771} (2017) 9--12, [\href{http://arxiv.org/abs/1609.08168}{{\tt arXiv:1609.08168}}].

\bibitem{LISACosmologyWorkingGroup:2022kbp}
{\bf LISA Cosmology Working Group} Collaboration, N.~Bartolo et~al., {\it {Probing anisotropies of the Stochastic Gravitational Wave Background with LISA}},  {\em JCAP} {\bf 11} (2022) 009, [\href{http://arxiv.org/abs/2201.08782}{{\tt arXiv:2201.08782}}].

\bibitem{Bartolo:2019oiq}
N.~Bartolo, D.~Bertacca, S.~Matarrese, M.~Peloso, A.~Ricciardone, A.~Riotto, and G.~Tasinato, {\it {Anisotropies and non-Gaussianity of the Cosmological Gravitational Wave Background}},  {\em Phys. Rev. D} {\bf 100} (2019), no.~12 121501, [\href{http://arxiv.org/abs/1908.00527}{{\tt arXiv:1908.00527}}].

\bibitem{Bartolo:2019yeu}
N.~Bartolo, D.~Bertacca, S.~Matarrese, M.~Peloso, A.~Ricciardone, A.~Riotto, and G.~Tasinato, {\it {Characterizing the cosmological gravitational wave background: Anisotropies and non-Gaussianity}},  {\em Phys. Rev. D} {\bf 102} (2020), no.~2 023527, [\href{http://arxiv.org/abs/1912.09433}{{\tt arXiv:1912.09433}}].

\bibitem{Ricciardone:2021kel}
A.~Ricciardone, L.~V. Dall'Armi, N.~Bartolo, D.~Bertacca, M.~Liguori, and S.~Matarrese, {\it {Cross-Correlating Astrophysical and Cosmological Gravitational Wave Backgrounds with the Cosmic Microwave Background}},  {\em Phys. Rev. Lett.} {\bf 127} (2021), no.~27 271301, [\href{http://arxiv.org/abs/2106.02591}{{\tt arXiv:2106.02591}}].

\bibitem{LISACosmologyWorkingGroup:2022jok}
{\bf LISA Cosmology Working Group} Collaboration, P.~Auclair et~al., {\it {Cosmology with the Laser Interferometer Space Antenna}},  {\em Living Rev. Rel.} {\bf 26} (2023), no.~1 5, [\href{http://arxiv.org/abs/2204.05434}{{\tt arXiv:2204.05434}}].

\bibitem{Bertacca:2019fnt}
D.~Bertacca, A.~Ricciardone, N.~Bellomo, A.~C. Jenkins, S.~Matarrese, A.~Raccanelli, T.~Regimbau, and M.~Sakellariadou, {\it {Projection effects on the observed angular spectrum of the astrophysical stochastic gravitational wave background}},  {\em Phys. Rev. D} {\bf 101} (2020), no.~10 103513, [\href{http://arxiv.org/abs/1909.11627}{{\tt arXiv:1909.11627}}].

\bibitem{Schulze:2023ich}
F.~Schulze, L.~Valbusa~Dall'Armi, J.~Lesgourgues, A.~Ricciardone, N.~Bartolo, D.~Bertacca, C.~Fidler, and S.~Matarrese, {\it {GW\_CLASS: Cosmological Gravitational Wave Background in the cosmic linear anisotropy solving system}},  {\em JCAP} {\bf 10} (2023) 025, [\href{http://arxiv.org/abs/2305.01602}{{\tt arXiv:2305.01602}}].

\bibitem{Malhotra:2022ply}
A.~Malhotra, E.~Dimastrogiovanni, G.~Dom\`enech, M.~Fasiello, and G.~Tasinato, {\it {New universal property of cosmological gravitational wave anisotropies}},  {\em Phys. Rev. D} {\bf 107} (2023), no.~10 103502, [\href{http://arxiv.org/abs/2212.10316}{{\tt arXiv:2212.10316}}].

\bibitem{ValbusaDallArmi:2023nqn}
L.~Valbusa~Dall'Armi, A.~Mierna, S.~Matarrese, and A.~Ricciardone, {\it {Adiabatic or Non-Adiabatic? Unraveling the Nature of Initial Conditions in the Cosmological Gravitational Wave Background}},  \href{http://arxiv.org/abs/2307.11043}{{\tt arXiv:2307.11043}}.

\bibitem{Isaacson1}
R.~A. Isaacson, {\it Gravitational radiation in the limit of high frequency. i. the linear approximation and geometrical optics},  {\em Phys. Rev.} {\bf 166} (Feb, 1968) 1263--1271.

\bibitem{Isaacson2}
R.~A. Isaacson, {\it Gravitational radiation in the limit of high frequency. ii. nonlinear terms and the effective stress tensor},  {\em Phys. Rev.} {\bf 166} (Feb, 1968) 1272--1280.

\bibitem{Pizzuti:2022nnj}
L.~Pizzuti, A.~Tomella, C.~Carbone, M.~Calabrese, and C.~Baccigalupi, {\it {Boltzmann equations for astrophysical Stochastic Gravitational Wave Backgrounds scattering off of massive objects}},  {\em JCAP} {\bf 02} (2023) 054, [\href{http://arxiv.org/abs/2208.02800}{{\tt arXiv:2208.02800}}].

\bibitem{Cusin:2018avf}
G.~Cusin, R.~Durrer, and P.~G. Ferreira, {\it {Polarization of a stochastic gravitational wave background through diffusion by massive structures}},  {\em Phys. Rev. D} {\bf 99} (2019), no.~2 023534, [\href{http://arxiv.org/abs/1807.10620}{{\tt arXiv:1807.10620}}].

\bibitem{Garoffolo:2022usx}
A.~Garoffolo, {\it {Wave-optics limit of the stochastic gravitational wave background}},  {\em Phys. Dark Univ.} {\bf 44} (2024) 101475, [\href{http://arxiv.org/abs/2210.05718}{{\tt arXiv:2210.05718}}].

\bibitem{Dalang:2021qhu}
C.~Dalang, G.~Cusin, and M.~Lagos, {\it {Polarization distortions of lensed gravitational waves}},  {\em Phys. Rev. D} {\bf 105} (2022), no.~2 024005, [\href{http://arxiv.org/abs/2104.10119}{{\tt arXiv:2104.10119}}].

\bibitem{Cusin:2019rmt}
G.~Cusin and M.~Lagos, {\it {Gravitational wave propagation beyond geometric optics}},  {\em Phys. Rev. D} {\bf 101} (2020), no.~4 044041, [\href{http://arxiv.org/abs/1910.13326}{{\tt arXiv:1910.13326}}].

\bibitem{Pijnenburg:2024btj}
M.~Pijnenburg, G.~Cusin, C.~Pitrou, and J.-P. Uzan, {\it {Wave optics lensing of gravitational waves: theory and phenomenology of triple systems in the LISA band}},  \href{http://arxiv.org/abs/2404.07186}{{\tt arXiv:2404.07186}}.

\bibitem{Oancea:2022szu}
M.~A. Oancea, R.~Stiskalek, and M.~Zumalac\'arregui, {\it {Frequency- and polarization-dependent lensing of gravitational waves in strong gravitational fields}},  \href{http://arxiv.org/abs/2209.06459}{{\tt arXiv:2209.06459}}.

\bibitem{Oancea:2023hgu}
M.~A. Oancea, R.~Stiskalek, and M.~Zumalac\'arregui, {\it {Probing general relativistic spin-orbit coupling with gravitational waves from hierarchical triple systems}},  \href{http://arxiv.org/abs/2307.01903}{{\tt arXiv:2307.01903}}.

\bibitem{Chandrasekhar:1985kt}
S.~Chandrasekhar, {\em {The mathematical theory of black holes}}.
\newblock 1985.

\bibitem{Berti:2009kk}
E.~Berti, V.~Cardoso, and A.~O. Starinets, {\it {Quasinormal modes of black holes and black branes}},  {\em Class. Quant. Grav.} {\bf 26} (2009) 163001, [\href{http://arxiv.org/abs/0905.2975}{{\tt arXiv:0905.2975}}].

\bibitem{Konoplya:2011qq}
R.~A. Konoplya and A.~Zhidenko, {\it {Quasinormal modes of black holes: From astrophysics to string theory}},  {\em Rev. Mod. Phys.} {\bf 83} (2011) 793--836, [\href{http://arxiv.org/abs/1102.4014}{{\tt arXiv:1102.4014}}].

\bibitem{Hinderer:2007mb}
T.~Hinderer, {\it {Tidal Love numbers of neutron stars}},  {\em Astrophys. J.} {\bf 677} (2008) 1216--1220, [\href{http://arxiv.org/abs/0711.2420}{{\tt arXiv:0711.2420}}]. [Erratum: Astrophys.J. 697, 964 (2009)].

\bibitem{Binnington:2009bb}
T.~Binnington and E.~Poisson, {\it {Relativistic theory of tidal Love numbers}},  {\em Phys. Rev. D} {\bf 80} (2009) 084018, [\href{http://arxiv.org/abs/0906.1366}{{\tt arXiv:0906.1366}}].

\bibitem{Damour:2009vw}
T.~Damour and A.~Nagar, {\it {Relativistic tidal properties of neutron stars}},  {\em Phys. Rev. D} {\bf 80} (2009) 084035, [\href{http://arxiv.org/abs/0906.0096}{{\tt arXiv:0906.0096}}].

\bibitem{Pound:2021qin}
A.~Pound and B.~Wardell, {\it {Black hole perturbation theory and gravitational self-force}},  \href{http://arxiv.org/abs/2101.04592}{{\tt arXiv:2101.04592}}.

\bibitem{Teukolsky:1973ha}
S.~A. Teukolsky, {\it {Perturbations of a rotating black hole. 1. Fundamental equations for gravitational electromagnetic and neutrino field perturbations}},  {\em Astrophys. J.} {\bf 185} (1973) 635--647.

\bibitem{Newman:1961qr}
E.~Newman and R.~Penrose, {\it {An Approach to gravitational radiation by a method of spin coefficients}},  {\em J. Math. Phys.} {\bf 3} (1962) 566--578.

\bibitem{Gloge1969FormalQT}
D.~C. Gloge and D.~Marcuse, {\it Formal quantum theory of light rays},  {\em Journal of the Optical Society of America} {\bf 59} (1969) 1629--1631.

\bibitem{Stoler:1981}
D.~{Stoler}, {\it {Operator methods in physical optics}},  {\em Journal of the Optical Society of America (1917-1983)} {\bf 71} (Mar., 1981) 334.

\bibitem{guralnik2019new}
Z.~Guralnik, {\it A new look at the helmholtz equation: Lefschetz thimbles and the einbein action},  2019.

\bibitem{Feynman:100771}
R.~P. Feynman and A.~R. Hibbs, {\em {Quantum mechanics and path integrals}}.
\newblock International series in pure and applied physics. McGraw-Hill, New York, NY, 1965.

\bibitem{feldbrugge2020gravitational}
J.~Feldbrugge and N.~Turok, {\it Gravitational lensing of binary systems in wave optics},  2020.

\bibitem{Tambalo_2023}
G.~Tambalo, M.~Zumalacárregui, L.~Dai, and M.~H.-Y. Cheung, {\it Lensing of gravitational waves: Efficient wave-optics methods and validation with symmetric lenses},  {\em Physical Review D} {\bf 108} (Aug., 2023).

\bibitem{cheung2024probing}
M.~H.-Y. Cheung, K.~K.~Y. Ng, M.~Zumalacárregui, and E.~Berti, {\it Probing minihalo lenses with diffracted gravitational waves},  2024.

\bibitem{PhysRevD.107.043029}
M.~\ifmmode \mbox{\c{C}}\else \c{C}\fi{}al\ensuremath{\iota}\ifmmode~\mbox{\c{s}}\else \c{s}\fi{}kan, L.~Ji, R.~Cotesta, E.~Berti, M.~Kamionkowski, and S.~Marsat, {\it Observability of lensing of gravitational waves from massive black hole binaries with lisa},  {\em Phys. Rev. D} {\bf 107} (Feb, 2023) 043029.

\bibitem{Jow:2022pux}
D.~L. Jow, U.-L. Pen, and J.~Feldbrugge, {\it {Regimes in astrophysical lensing: refractive optics, diffractive optics, and the Fresnel scale}},  {\em Mon. Not. Roy. Astron. Soc.} {\bf 525} (2023), no.~2 2107--2124, [\href{http://arxiv.org/abs/2204.12004}{{\tt arXiv:2204.12004}}].

\bibitem{thorne2021optics}
K.~Thorne and R.~Blandford, {\em Optics: Volume 2 of Modern Classical Physics}.
\newblock Princeton University Press, 2021.

\bibitem{Garrod_1966}
C.~GARROD, {\it Hamiltonian path-integral methods},  {\em Rev. Mod. Phys.} {\bf 38} (Jul, 1966) 483--494.

\bibitem{Feldbrugge:2023frq}
J.~Feldbrugge, D.~L. Jow, and U.-L. Pen, {\it {Complex classical paths in quantum reflections and tunneling}},  \href{http://arxiv.org/abs/2309.12420}{{\tt arXiv:2309.12420}}.

\bibitem{DeWittMorette1979PathII}
C.~DeWitt-Morette, A.~N. Maheshwari, and B.~Nelson, {\it Path integration in non-relativistic quantum mechanics},  {\em Physics Reports} {\bf 50} (1979) 255--372.

\bibitem{ThorneBlandford}
K.~S. {Thorne} and R.~D. {Blandford}, {\em {Modern Classical Physics: Optics, Fluids, Plasmas, Elasticity, Relativity, and Statistical Physics}}.
\newblock 2017.

\bibitem{Pfenning:2000zf}
M.~J. Pfenning and E.~Poisson, {\it {Scalar, electromagnetic, and gravitational selfforces in weakly curved space-times}},  {\em Phys. Rev. D} {\bf 65} (2002) 084001, [\href{http://arxiv.org/abs/gr-qc/0012057}{{\tt gr-qc/0012057}}].

\bibitem{Chu:2019vmh}
Y.-Z. Chu, K.~Pasmatsiou, and G.~D. Starkman, {\it {Finite-size effects on the self-force}},  {\em Phys. Rev. D} {\bf 101} (2020), no.~10 104020, [\href{http://arxiv.org/abs/1910.02924}{{\tt arXiv:1910.02924}}].

\bibitem{Copi:2022ire}
C.~Copi and G.~D. Starkman, {\it {Gravitational Glint: Detectable Gravitational Wave Tails from Stars and Compact Objects}},  {\em Phys. Rev. Lett.} {\bf 128} (2022), no.~25 251101, [\href{http://arxiv.org/abs/2201.03684}{{\tt arXiv:2201.03684}}].

\bibitem{Chu:2011ip}
Y.-Z. Chu and G.~D. Starkman, {\it {Retarded Green's Functions In Perturbed Spacetimes For Cosmology and Gravitational Physics}},  {\em Phys. Rev. D} {\bf 84} (2011) 124020, [\href{http://arxiv.org/abs/1108.1825}{{\tt arXiv:1108.1825}}].

\bibitem{Peskin:1995ev}
M.~E. Peskin and D.~V. Schroeder, {\em {An Introduction to quantum field theory}}.
\newblock Addison-Wesley, Reading, USA, 1995.

\bibitem{Fetter}
A.~L. Fetter and J.~D. Walecka, {\em Quantum Theory of Many-Particle Systems}.
\newblock McGraw-Hill, Boston, 1971.

\bibitem{Schwartz:2014sze}
M.~D. Schwartz, {\em {Quantum Field Theory and the Standard Model}}.
\newblock Cambridge University Press, 3, 2014.

\bibitem{Srednicki:2007qs}
M.~Srednicki, {\em {Quantum field theory}}.
\newblock Cambridge University Press, 1, 2007.

\bibitem{Bertacca:2017vod}
D.~Bertacca, A.~Raccanelli, N.~Bartolo, and S.~Matarrese, {\it {Cosmological perturbation effects on gravitational-wave luminosity distance estimates}},  {\em Phys. Dark Univ.} {\bf 20} (2018) 32--40, [\href{http://arxiv.org/abs/1702.01750}{{\tt arXiv:1702.01750}}].

\bibitem{Tambalo:2022plm}
G.~Tambalo, M.~Zumalac\'arregui, L.~Dai, and M.~H.-Y. Cheung, {\it {Lensing of gravitational waves: Efficient wave-optics methods and validation with symmetric lenses}},  {\em Phys. Rev. D} {\bf 108} (2023), no.~4 043527, [\href{http://arxiv.org/abs/2210.05658}{{\tt arXiv:2210.05658}}].

\bibitem{Morita:2019sau}
T.~Morita and J.~Soda, {\it {Arrival Time Differences of Lensed Massive Gravitational Waves}},  \href{http://arxiv.org/abs/1911.07435}{{\tt arXiv:1911.07435}}.

\bibitem{Regge:1957td}
T.~Regge and J.~A. Wheeler, {\it {Stability of a Schwarzschild singularity}},  {\em Phys. Rev.} {\bf 108} (1957) 1063--1069.

\bibitem{Zerilli:1970se}
F.~J. Zerilli, {\it {Effective potential for even parity Regge-Wheeler gravitational perturbation equations}},  {\em Phys. Rev. Lett.} {\bf 24} (1970) 737--738.

\bibitem{Maggiore:2018sht}
M.~Maggiore, {\em {Gravitational Waves. Vol. 2: Astrophysics and Cosmology}}.
\newblock Oxford University Press, 3, 2018.

\bibitem{Press:1973zz}
W.~H. Press and S.~A. Teukolsky, {\it {Perturbations of a Rotating Black Hole. II. Dynamical Stability of the Kerr Metric}},  {\em Astrophys. J.} {\bf 185} (1973) 649--674.

\bibitem{Andersson:2020gsj}
L.~Andersson, J.~Joudioux, M.~A. Oancea, and A.~Raj, {\it {Propagation of polarized gravitational waves}},  {\em Phys. Rev. D} {\bf 103} (2021), no.~4 044053, [\href{http://arxiv.org/abs/2012.08363}{{\tt arXiv:2012.08363}}].

\end{thebibliography}\endgroup
\bibliographystyle{JHEP}

\end{document}